\documentclass{jfm}
\usepackage{multirow}
\usepackage[table]{xcolor}
\usepackage{rotating}
\usepackage{amsmath}
\usepackage{bm}
\usepackage{array}
\usepackage{graphicx}
\usepackage{dcolumn}   
\usepackage{amssymb}
\usepackage[english]{babel}
\usepackage[colorlinks=true, citecolor=red, linkcolor=blue,urlcolor=blue ]{hyperref}
\usepackage{epstopdf, epsfig}

\shorttitle{General geometric phase methods}
\shortauthor{L. Koens and E. Lauga}

\title{Geometric phase methods with Stokes theorem for a general viscous swimmer}

\author{Lyndon Koens\aff{1}
  \corresp{\email{lmk42@cam.ac.uk}}
 \and  Eric Lauga\aff{2}}
\affiliation{\aff{1}Department of Mathematics and Statistics, Macquarie University, 192 Balaclava Rd, Macquarie Park, NSW 2113, Australia
\aff{2}Department of Applied Mathematics and Theoretical Physics, University of Cambridge, Wilberforce Road, Cambridge CB3 0WA, United Kingdom}

\begin{document}

\maketitle

\begin{abstract}
The geometric phase techniques for swimming in viscous flows  express the net displacement of a swimmer as a path integral of a field in configuration space. This representation can be transformed into an area integral for simple swimmers using Stokes theorem. Since this transformation applies for any loop, the integrand of this area integral can be used to help design these swimmers. However, the extension of this Stokes theorem technique to more complicated swimmers is hampered by problems with variables that do not commute and by how to visualise and understand the higher dimensional spaces. In this paper, we develop a treatment for each of these problems, thereby allowing the displacement of general swimmers in any environment to be designed and understood similarly to simple swimmers. The net displacement arising from non-commuting variables is tackled by embedding the integral into a higher dimensional space, which can then be visualised through a suitability constructed surface.  These methods are developed for general swimmers and demonstrated on {three} benchmark examples: Purcell's two-hinged swimmer, an axisymmetric squirmer in free space {and an axisymmetric squirmer approaching a free interface}.  We show in particular that for swimmers with more than two modes of deformation, there exists an infinite set of strokes that generate each net displacement. Hence, in the absence of additional restrictions,  general microscopic swimmers do not have a single stroke that maximises their displacement. 
\end{abstract}

\section{Introduction}

The success of \citet{GRAY1955} resistive-force model for spermatozoa swimming  inspired a great effort into the modelling of microscopic biological swimmers in viscous fluids. Hydrodynamic models of these systems   now include long-range hydrodynamic interactions in flagella and cilia \citep{1976, Keller1976a, Johnson1979,Gueron1992, Man2016, Koens2018}, interactions between the swimmers body and its flagella \citep{Higdon2006,Smith2009a,Hu2015,Chakrabarti2019a}, and with walls \citep{Barta1988, Walker2018,Das2018}. These hydrodynamic models have also been coupled with elasticity to investigate the fluid-structure interactions at the heart of biological flagella \citep{Kim2005b, Ishimoto2018, Chakrabarti2019b, DuRoure2019}, while including  models for internal activation allows us to understand how spermatozoa and cilia generate their waveforms  \citep{Chakrabarti2019, Chakrabarti2019a, Man2020, Ishimoto2018}. These approaches  have also been extended to consider the dynamics of microswimmers in complex media \citep{Omori2019, Hewitt2018, Koens2016a, Wrobel2016}. 
These theoretical developments, carried out in close collaboration with experiments \citep{Martinez2020, Turner2000, Bianchi2017, Drescher2010, Goldstein2015, PerezIpina2019, Colin2019}, have lead to an increased understandings of how bacteria \citep{Lauga2016}, algae \citep{Goldstein2015} and spermatozoa \citep{Gaffney2011} interact with their environment  and with each other and prompted the creation of artificial microscopic swimmers to test hydrodynamic theories \citep{Valdes2019,Hayashi2020}, explore collective active systems \citep{Karani2019,Alapan2019}, and for the development of micro-technologies such as targeted drug delivery \citep{Maggi2015b, Huang2019, Zhang2010, Vizsnyiczai2017, Koens2019}.

The derivation of these theoretical models have occurred in conjunction with important theoretical developments in our fundamental understanding of the world of viscous flows. For example, in 1951, Taylor first determined the mathematics of swimming at low Reynolds number and showed that it non-linearly depended on the waving geometry \citep{Taylor1951}. This paved the way for the work of \citet{GRAY1955}. \citet{Purcell} later argued that bodies could not swim in viscous fluids unless they broke the time-symmetry of the system (\textit{i.e.} formed a so-called non-reciprocal stroke), which he demonstrated with his simple two-hinged swimmer. The motion of Purcell's swimmer was not solved until much later \citep{BECKER2003} but has since been studied extensively \citep{Gutman2016, Wiezel2018, Avron2008, Hatton2015, Ramasamy2016,Ramasamy2017, Ramasamy2019, Hatton2011, Hatton2013} and  been extended to create other prototypical low Reynolds number swimmers \citep{Golestanian2008}. \citet{1976} also used geometric arguments to   determine asymptotically the flow around slender filaments such as those used by bacteria and spermatozoa. This has lead to models of filaments in other geometries and with other shapes \citep{Borker2019,Koens2016}.

One significant, but far less used, theoretical development is the geometric swimming formulation due to \citet{Shapere1987,Shapere1989} who recognised that, as the displacement of any viscous swimmer is purely a function of geometry, it could be represented by a gauge theory. These gauge theories have been studied at length in physics and so this representation introduces many new ideas and techniques. For swimmers that travel in one dimension and only have two degrees of freedom, the geometrical techniques could be further simplified through the use of the Stokes theorem.  This simplification allows the net displacement from any stroke to be visualised on a plane and thereby provides new insights into the propulsion mechanisms and how to  design the swimmer's displacement \citep{Koens2018a, Shapere1989, Desimone2012, DeSimone2012a, Cicconofri2016,Hatton2015}. However the same treatment is not generally possible for more complicated swimmers because of (i) non-commuting variables within the gauge field and (ii) difficulties visualising the configuration space \citep{Hatton2015,Hatton2017,Ramasamy2017, Bittner2018}. Attempts to produce similar results often restrict the motion of the swimmer and the types of deformation.

In this paper we show that the displacement of a microscopic swimmer in any environment can always be visualised on a single surface. This is achieved by developing methods to overcome both the non-commuting variables and visualisation issues for a general geometric swimming formulation. Our techniques are first described in the most general framework and then demonstrated with examples of prototypical microswimmers. The issue with non-commuting variables is overcome by embedding the system into higher dimensional spaces in which the Stokes theorem can again directly apply. The multidimensional equivalent of conversation of flux then allows us to visualise the dynamics in high-dimensional configuration spaces through a suitable surface. Significantly, our generalisation shows that for any swimmer in an unbound deformation space, there is no single stroke that maximises the displacement. Finally we outline how the techniques could be used to choose the displacement of an arbitrary  unconstrained viscous swimmer.

The paper is organised as follows. Section~\ref{sec:low}  summaries the geometric swimming formalism for viscous swimmers and explains the difficulties posed with complex swimmers and some of the attempts to get around them. This development and associated difficulty is demonstrated with the rotation of a small-angled Purcell's two-hinged swimmer. Section~\ref{sec:commute} then introduces a method to overcome non-commutating variables so that the generalized Stokes theorem can be applied. This is demonstrated in the case of the translation of the aforementioned Purcell swimmer. In Sec.~\ref{sec:vis} we next develop a method to understand and visualise the full space configuration space after the use of Stokes theorem. We show in particular that the whole space can be viewed on a single surface and that the net displacement is not unique to any one stroke, as demonstrated with a four-mode squirmer { and the translation of the Purcell swimmer}.
 Finally a general process to design a swimmer's displacement is outlined in Sec.~\ref{sec:design}, {demonstrated with a squirmer near an interface,} and our results are summarised in Sec.~\ref{sec:con}.

\section{Background: Geometric swimming for Stokes flow} \label{sec:low}

The geometric swimming techniques for Stokes flow are  natural consequences of the linearity and time independence of viscous flow. In this section we outline how the geometric swimming representation comes about, how the Stokes theorem can be used for simple swimmers and the difficulty with this extension in more complicated set-ups.

\subsection{Origin of the geometric swimming representation} \label{sec:dis1}

Microscopic swimmers typically exist within inertia-less environments \citep{Lauga2009}. As a result the swimmers are force and torque-free and the surrounding fluid flow is well described by the incompressible Stokes equations,
\begin{eqnarray}
\mu \nabla^{2} \mathbf{u} &=&  \nabla p, \\
\nabla \cdot \mathbf{u} &=& 0, 
\end{eqnarray}
where $\mu$ is the dynamic viscosity, $\mathbf{u}$ is the fluid velocity, and $p$ is the dynamic pressure. The typical boundary conditions for these equations are the no-slip condition on the swimmers surface, \textit{i.e.} the fluid velocity must equal the velocity on the swimmers surface. The linear and time independence of the above equations mean that the swimming problem can always be broken into two parts: the response to surface deformation and the response to rigid body motion. The total motion is then found by adding the two flows together afterwards. In each of these cases the forces and torques must be linearly related to an appropriate velocity. The hydrodynamic force, $\mathbf{F}_{R}$, and torque, $\mathbf{L}_{R}$, on the swimmer from rigid body motion can therefore be written as
\begin{equation}
\left(\begin{array}{c}
\mathbf{F}_{R} \\
\mathbf{L}_{R}
\end{array} \right) = - \mathbi{R}(\mathbf{l}, \mathbf{x}) \cdot\left(\begin{array}{c}
\mathbf{U} \\
\boldsymbol{\Omega}
\end{array} \right), \label{RM}
\end{equation}
where $\mathbf{U}$ is the rigid body velocity, $\boldsymbol{\Omega}$ is the rigid body angular velocity, and $\mathbi{R}(\mathbf{l}, \mathbf{x})$ is the resistance matrix. The resistance matrix is only a function of the swimmer's configuration and laboratory frame position and orientation. In the above we have assumed that the swimmer's configuration is uniquely described by a set of deformation modes, $\mathbf{l}$, while the position and orientation in the laboratory frame is given by $\mathbf{x}$. Similarly the hydrodynamic force, $\mathbf{F}_{d}$, and torque, $\mathbf{L}_{d}$, on the body from the shape deformation can be written as
\begin{equation}
\left(\begin{array}{c}
\mathbf{F}_{d} \\
\mathbf{L}_{d}
\end{array} \right) = \mathbi{A}(\mathbf{l}, \mathbf{x}) \cdot\frac{d \mathbf{l}}{d t},
\end{equation}
where  $\mathbi{A}(\mathbf{l}, \mathbf{x})$ is also a matrix which depends only on the swimmers configuration and position. The matrices $\mathbi{R}(\mathbf{l}, \mathbf{x})$ and $\mathbi{A}(\mathbf{l}, \mathbf{x})$ must be determined by solving the Stokes equations with the appropriate boundary conditions. Except in systems with suitable symmetries, such as  spheroids, this is difficult  to do and many analytical and computational techniques have been developed to tackle  this \citep{Pozrikidis1992,Cortez2005,Cortez2001,1976,Keller1976a, Johnson1979,Kim2005}. 

If these matrices can be found, the instantaneous velocity of a swimmer can be determined by balancing the forces and torques on the swimmer. Since low Reynolds number swimmers are force and torque free, their velocities can be written as
\begin{equation}
\left(\begin{array}{c}
\mathbf{U} \\
\boldsymbol{\Omega}
\end{array} \right) = \mathbi{R}^{-1}(\mathbf{l}, \mathbf{x})\cdot \mathbi{A}(\mathbf{l}, \mathbf{x})\cdot \frac{d \mathbf{l}}{d t}, \label{guage}
\end{equation}
 in the absence of external forces or torques. Note that the inversion of $\mathbf{R}$ is always possible as the resistance tensor is positive definite \citep{Kim2005}.
The evolution of the swimmers position and orientation in the laboratory frame is therefore given by
\begin{equation}
\frac{d \mathbf{x}}{dt} = \mathbi{B}(\mathbf{x})\cdot\left(\begin{array}{c}
\mathbf{U} \\
\boldsymbol{\Omega}
\end{array} \right) = \mathbi{M}(\mathbf{l}, \mathbf{x}) \cdot \frac{d \mathbf{l}}{d t} \label{evo}
\end{equation}
where $\mathbi{B}(\mathbf{x})$ is the evolution matrix which relates the swimmers velocities to the rate of change of its coordinates. Of course, $\mathbi{B}(\mathbf{x})$ is different for different representations of $\mathbf{x}$. In the above $\mathbi{M}(\mathbf{l}, \mathbf{x})$ is the gauge field identified by \citet{Shapere1987,Shapere1989}.

The position of the swimmer after a specific deformation $\mathbf{l}(t)$ is given by integrating the evolution equation, Eq.~\eqref{evo}, over this path. Mathematically this can be written as
\begin{equation}
\mathbf{x}(t) = \int_{0}^{t} \mathbi{M}(\mathbf{l}(t'), \mathbf{x}(t')) \cdot \frac{d \mathbf{l}}{d t'}\,d t'. \label{pos}
\end{equation}
The above equation is therefore a parametrised path integral for the path $\mathbf{l}(t)$. This holds for a general path and so any displacement is given by a path integral over the field $\mathbi{M}(\mathbf{l}, \mathbf{x})$. The net displacement from a periodic swimming stroke can thus always be written as the  path integral
\begin{equation}
\Delta\mathbf{x} = \oint_{\partial V} \mathbi{M}(\mathbf{l}, \mathbf{x})\cdot d \mathbf{l}, \label{net}
\end{equation}
where $\partial V$ is the prescribed loop in the configuration space. Hence $\mathbi{M}(\mathbf{l}, \mathbf{x})$ contains all the information about the swimmers displacement. This representation can be used used to  design  the swimmer's displacement by identifying how $\Delta \mathbf{x}$ changes over many different strokes.

\subsection{Extensions for simple swimmers}

{ The analysis of the geometric swimming results can be made easier} for isolated one-dimensional swimmers with two degrees of freedom, $\mathbf{l} =\left\{l_{1},l_{2}\right\}$. In this case $\mathbi{M}(\mathbf{l}, \mathbf{x})=\left\{M_{1}(l_{1},l_{2}),M_{2}(l_{1},l_{2}) \right\}$ is independent of the laboratory frame configuration and the translational displacement can be written as
\begin{eqnarray}
\Delta x &=& \oint_{\partial V}\left(M_{1}(l_{1},l_{2}) \frac{d l_{1}}{dt} + M_{2}(l_{1},l_{2})\frac{d l_{2}}{dt} \right) \,d t \notag \\
&=&-\iint_{V}\left(\frac{\partial M_{1}}{\partial l_{2}} -  \frac{\partial M_{2}}{\partial l_{1}} \right) \,dl_{1} \,dl_{2}
\end{eqnarray}
where we have used used Green's theorem (i.e.~the two-dimensional version of Stokes theorem) and where $V$ is the area inside the loop. This representation applies to any choice of loop, and hence the displacement of the swimmer is inherently related to $-\left(\displaystyle \frac{\partial M_{1}}{\partial l_{2}} -  \frac{\partial M_{2}}{\partial l_{1}} \right)$. Inspection of this function on the $\left\{l_{1},l_{2}\right\}$ plane therefore displays the general behaviour of the displacement in an easy-to-visualise way.  This can be useful for the design and selection of swimming strokes for specific tasks. This representation is also  attractive from a theoretical perspective because swimming strokes that do not break time-reversal symmetry contain zero area within them and so clearly produce no displacement. Such is the appeal of this representation that it has been used popularly under the names motility maps \citep{Koens2018a, Desimone2012, DeSimone2012a, Cicconofri2016} and  height functions (See Ref.~\citet{Hatton2011} and references within).

\subsection{Difficulties in the extension to more complicated swimmers}

The extension of the above Green's theorem idea to the general geometric swimmer method encounters two difficulties: (i) the visualisation of the results and (ii) the treatment of non-commuting variables.  

First, the visualisation issue can be seen if we consider a field $\mathbi{M}(\mathbf{l}, \mathbf{x})$ which is independent of  $\mathbf{x}$. In this case the generalized Stokes theorem,
\begin{equation}
\int_{\partial V} w =\int_{ V} dw, 
\end{equation}
 can be used to relate the closed line integral to the flux of a field through surfaces bounded by the loop in any dimensional space. The above equation uses notation from exterior calculus and so a summary of the relevant features has been included in Appendix~\ref{sec:extior}. The generalized Stokes theorem allows the displacement to be written as
 \begin{eqnarray}
 \Delta x_{i} &=& \oint_{\partial V}M_{ij}  d l^{j} \notag \\
 &=& \iint_{ V} d\left(M_{ij} d l^{j}\right) \notag \\
 &=& -\frac{1}{2}  \iint_{V}  \left( \frac{\partial M_{ij}}{\partial l^{k}} -\frac{\partial M_{ik}}{\partial l^{j}}  \right) d l^{j} \wedge d l^{k} \label{displacement}
 \end{eqnarray}
where $\wedge$ is the wedge product (the antisymmetric tensor product), $d$ is the exterior derivative, and $V$ is any surface bounded by the loop $\partial V$. In the above we have used index notation, the Einstein summation convention, and have invoked $d l^{j} \wedge d l^{i} = - d l^{i} \wedge d l^{j}$ and $d^{2} w =0$ for any $w$. Physically $d l^{j} \wedge d l^{k}$ represents the infinitesimal surface element between directions $l^{j}$ and $l^{k}$ and, for a specific direction $i$, $\partial M_{jk}^{i}=\displaystyle  -\frac{1}{2}\left( \frac{\partial M_{ij}}{\partial l^{k}} -\frac{\partial M_{ik}}{\partial l^{j}}  \right)$ represents a skew-symmetric matrix of size $N\times N$ where $N$ is the size of $\mathbf{l}$. This skew-symmetric matrix can be thought as the multidimensional equivalent to the curl (Appendix~\ref{sec:extior}). Similarly to Stokes theorem in three dimensions, this integral can be interpreted as the flux of the multidimensional curl through the surface $V$. The visualisation of this field is however non-trivial for any swimmer with $N>3$, as it is not possible to plot the full field in a simple way. The optimization and design of general swimmers requires therefore the development of techniques to efficiently search these high dimensional spaces \citep{Ramasamy2017, Ramasamy2019}.

Secondly, the dependence of $\mathbi{M}(\mathbf{l}, \mathbf{x})$ on the laboratory position and orientation further complicates the extension of Green's theorem to other swimmers. This complication even occurs for two-dimensional swimmers in unbound fluids because rotations and translations do not commute. For example the marching orders ``Step, Turn", creates a different result to ``Turn, Step". Under these conditions the generalized Stokes theorem needs to be corrected to account for the change of $\mathbf{x}$ along the path. These corrections can be written as an infinite series of nested Lie brackets which are closely related to the Magnus expansion (a generalisation of the matrix exponential for non-commuting matrix equations) \citep{Hatton2015, Ramasamy2017, Bittner2018, Radford,Ramasamy2019}. This infinite series can be interpreted as expansion around small loop sizes in which $\mathbf{x}$ changes little, because each Lie bracket captures the change produced by the non-commuting variables from an infinitesimal loop in the configuration space. Nested Lie brackets therefore capture the change of a change and so are more important when $\mathbf{x}$ varies more. This series cannot typically be computed, and so is often truncated to leading order in loop size \citep{Shapere1987,Shapere1989}. The leading correction in this series goes as
\begin{eqnarray}
\Delta \mathbf{x} = \oint_{\partial V} \mathbi{M}\cdot d \mathbf{l} & =&    \iint_{V} d(\mathbi{M}\cdot d \mathbf{l} )  + \sum_{i<j} \iint_{V} \left[\mathbf{M}_{i},\mathbf{M}_{j} \right] dl^{i} \wedge dl^{j}  \notag \\
 &&+ O\left(\iiint \left[\left[\mathbf{M}_{i},\mathbf{M}_{j} \right],\mathbf{M}_{k}\right]  dl^{i} \wedge dl^{j} \wedge dl^{k}\right) \label{Lie}
\end{eqnarray}
where $\mathbf{M}_{i}$ is the the $i$th column of $\mathbi{M}$, and $\left[\mathbf{X},\mathbf{Y} \right] = \mathbf{X} \cdot \nabla \mathbf{Y} - \mathbf{Y} \cdot \nabla \mathbf{X}$ is the Lie bracket of the two vector fields \citep{Radford}. { Each subsequent term in this expansion involves an additional integral over the configuration loop. Hence, if $\ell$ is the typical size of the stroke, the $n$th term in the expansion scales with $\ell^n$, reflecting that the series representation applies in the small $\ell$ limit.}

{ This limit means that the error on this representation can become very large. Consider for example the one dimensional field $\mathbf{M} =\{a, b x\}$, where $a$ and $b$ are constants. This field is similar to that of an axisymmetric squirmer approaching a free interface (Sec.~\ref{sec:squirmint}) and corresponds to the evolution equation
\begin{equation} 
\frac{d x}{d t} = a \frac{d l_{1}}{d t} + b  \frac{d l_{2}}{d t}  x(t), \label{demo}
\end{equation}
which can be solved exactly to find the net displacement
\begin{equation}
\Delta x = a e^{b l_2(0)} \int_{0}^{2\pi} e^{-b l_{2}(t)} \frac{d l_{1}}{d t} \,dt,
\end{equation}
where $t \in [0,2\pi)$. In this case Eq.~\eqref{Lie} predicts
\begin{equation}
\Delta x_{s} = 0-  a b \iint_{V} d l_{1} d l_{2} = -a b A,
\end{equation}
where $\Delta x_{s}$ is the predicted displacement from Eq.~\eqref{Lie}, and $A = \iint_{V} d l_{1} d l_{2} $ is the area enclosed in the loop. The series approximated displacement, $\Delta x_{s}$, is clearly very different to the exact solution $\Delta x$. This is because non-commuting variables often non-linearly affect the displacement while the first two terms in Eq.~\eqref{Lie} only account for the linear influence of these variables. As such $\Delta x_{s}$ is the leading order Taylor expansion of $\Delta x$ in small stroke size. 
}

{\citet{Hatton2011} noticed that the Lie brackets in the series expansion depends on the choice of frame. This meant that, though the error terms cannot be removed, their contribution can be minimised through an appropriate choice of frame, called the minimal perturbation frame. For isolated swimmers in an infinite fluid (i.e. when the matrices in Eq.~\eqref{guage} are independent of $\mathbf{x})$, they showed that this minimal perturbation frame can be determined through a generalised Helmholtz decomposition and solving a set of partial differential equations. This process cannot be done by hand in general and so the authors developed a numerical procedure to perform these calculations, available in MATLAB \citep{hatton_2020}. This numerical procedure limits the theoretical insight that can be developed from the technique but has been shown to improve significantly the results of the series expansion for two-dimensional Purcell-like low Reynolds number swimmers and some other two-dimensional non-low Reynolds number cases \citep{Hatton2011, Hatton2013, Hatton2015, Ramasamy2017, Hatton2017, Ramasamy2019, Bittner2018}. This process increases the range of applicability for the approximation, Eq.~\eqref{Lie}, for isolated swimmer problems. However no rigorous bound on the error after this improvement has been found and so the limits of this increase is not definitely known. {\color{blue} We note that in such two-dimensional isolated swimmer geometries, the displacement can be written as an exponential map \citep{Shapere1987,Shapere1989,Radford,Hatton2015}. Hence exponentiating the results of \citet{Hatton2011} further increases the effectiveness of this representation.}

\subsection{Example: A small-angled Purcell's two-hinged swimmer}

The power of these geometric techniques can be demonstrated using Purcell's famous two-hinged swimmer. Purcell's swimmer is considered one of the simplest model swimmers. It is composed of three rigid rods connected end-to-end. All the rods lie in a single plane and the angles between these rods can be varied (Fig.~\ref{fig:purcell}a). If the angles are changed in a non-reciprocal fashion this enables net motion within the two-dimensional plane \citep{Purcell}. \citet{BECKER2003} developed a detailed theoretical model for this swimmer. This model focused specifically on the swimming from the symmetric armed stroke proposed by Purcell in which one arm is raised, the second other aim is raised, the first is lowered and then the second is lowered. The motion and deformation of the swimmer was then determined by ensuring it was force and torque free and that the hinges imposed a specified torque difference across them. The motion of this swimmer has since been studied extensively in different scenarios \citep{Gutman2016, Wiezel2018, Avron2008, Hatton2015, Ramasamy2016, Ramasamy2017, Ramasamy2019, Hatton2011, Hatton2013}.

Unlike \citet{BECKER2003} work, we will specify the deformation and { typically } restrict ourselves to the limit of small angles between rods (assumed to be  of unit length). This makes the calculation tractable analytically for any stroke we wish to consider. { The full problem will be used to compare with the results of Eq.~\eqref{Lie} in the minimal perturbation coordinates.} 
 Here we specify the swimmers configuration, determine the gauge field $\mathbi{M}^{P}$ for the swimmer and demonstrate the simple extensions for the rotation of the swimmer. We also explain why this does not work directly for translation.

\subsubsection{Swimmer configuration and deformation velocity}

 \begin{figure}
\centering
\includegraphics[width=0.9\textwidth]{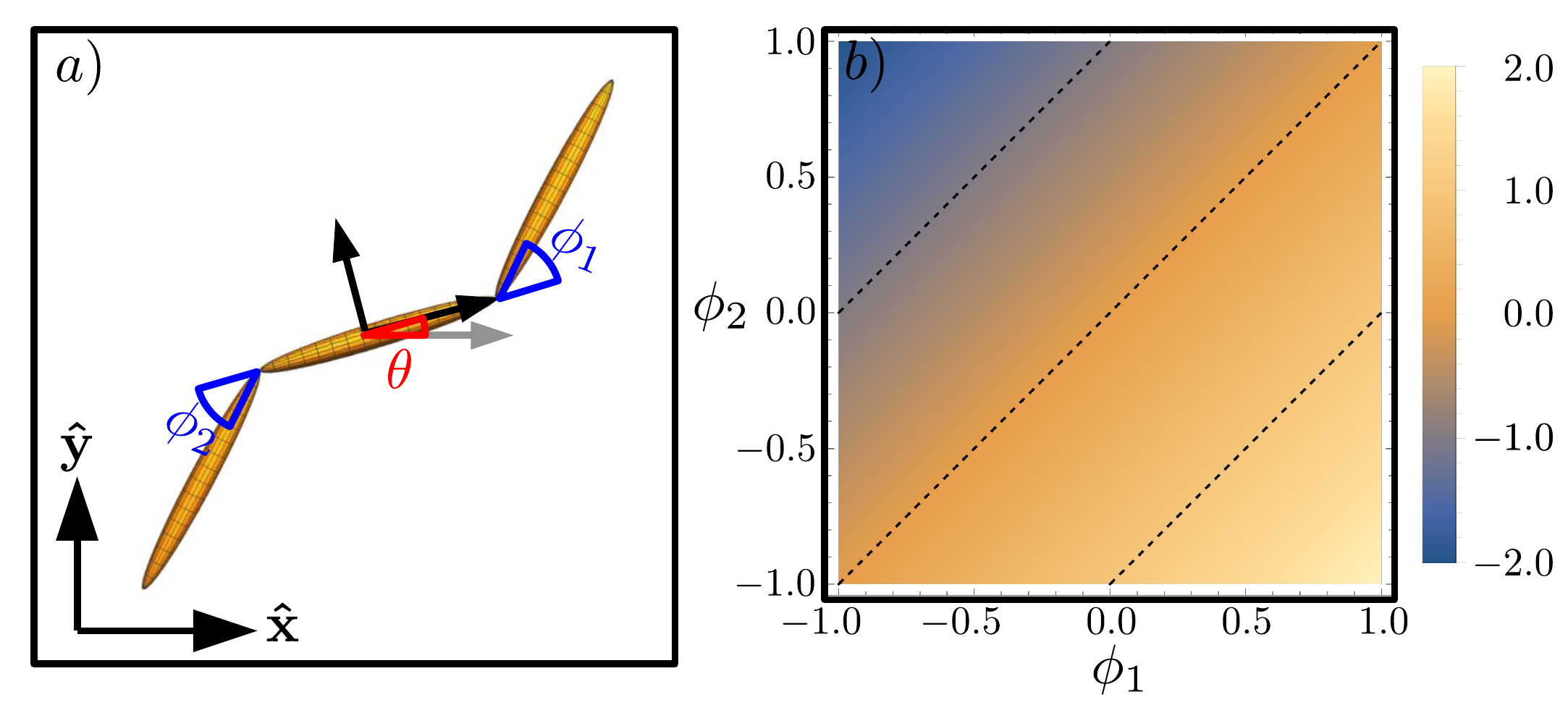}
\caption{a) Diagram of Purcell's two-hinged swimmer in the laboratory frame. The three slender rods are of unit length and the angle between the arms is $\phi_{1}$ and $\phi_{2}$. The laboratory orientation $\theta$ is defined with respect to the central rod. b) Iso-values of the integrand of Eq.~\eqref{rotG} where dashed lines represents contours. } 
\label{fig:purcell}
\end{figure} 

The configuration of Purcell's two-hinged swimmer and its deformation velocity are needed determine $\mathbi{M}^{P}$. If the component rods are slender, the structure of the swimmer can be described by the motion of the centreline of these rods, $\mathbf{r}(s,t)$. Hence, in a reference frame attached to the centre of the swimmer, its shape can be described by
\begin{equation}
\mathbf{r}(s,t)+\mathbf{r}_{c}(t) = \left\{\begin{array}{c r}
 \{-1+(1+s) \cos \phi_{1}(t),(1+s) \sin \phi_{1}(t)\} & -3<s<-1, \\
\{s ,0\} & -1<s<1,\\
 \{1+(s-1) \cos \phi_{2}(t),(s-1) \sin \phi_{2}(t)\} & 1<s<3,
\end{array} \right. 
\end{equation}
where $\mathbf{r}_{c}(t) = \{\cos\phi_{2}(t) -\cos\phi_{1}(t),\sin\phi_{2}(t) -\sin\phi_{1}(t)\}/3$ is the centre of swimmer at time $t$, $s$ is the arclength of the swimmer, and $\phi_{1}(t)$ and $\phi_{2}(t)$ are the angles between the first and second rod and the second and third rods respectively (Fig.~\ref{fig:purcell}a). This reference frame is a two-dimensional Cartesian coordinate frame $\{x',y'\}$ where $x'$ is aligned with the central rod called the body frame.
Similarly the surface velocity of the swimmer is approximately
\begin{equation}
\mathbf{V}(s,t) = \dot{\mathbf{r}}(s,t) = -\dot{\mathbf{r}}_{c}+\left\{\begin{array}{c r}
\dot{\phi_{1}} \{- (1+s) \sin \phi_{1},(1+s) \cos \phi_{1}\} & -3<s<-1, \\
\{0 ,0\} & -1<s<1,\\
\dot{\phi_{2}} \{-(s-1) \sin \phi_{2},(s-1) \cos \phi_{2}\} & 1<s<3, \label{def}
\end{array} \right. 
\end{equation}
where $\dot{(\cdot)}$ denotes the time derivative, we have dropped the time dependence for brevity and have neglected the thickness of the slender rods. This surface velocity has zero mean but can still exerts a non-zero net force and torque on the body, which   balance the force and torque from rigid body translation and rotation and thus generate motion.

\subsubsection{Forces and torques from deformation and rigid body motion}

The force and torques on Purcell's two-hinged swimmer can be estimated using resistive-force theory \citep{Lauga2009}. Resistive-force theory is an asymptotic result that captures the hydrodynamic force per unit length, $\mathbf{f}$, experienced by   a slender body in viscous flows in terms of the drag experienced if moved along its axis and perpendicular to it. Mathematically this relationship says
\begin{equation}
\mathbf{f} = -\left[\zeta_{\parallel} \mathbf{\hat{t}} \mathbf{\hat{t}} + \zeta_{\perp} \left( \mathbf{I}- \mathbf{\hat{t}} \mathbf{\hat{t}}\right) \right]\cdot\mathbf{U}, \label{RFT}
\end{equation}
where $\mathbf{U}$ is the velocity of the cylinder at $s$, $\mathbf{\hat{t}} = \partial_{s} \mathbf{r}$ is the local tangent vector of the swimmers body, $\zeta_{\parallel}$ is the drag coefficient  for motion along the axis and $\zeta_{\perp}$ is the drag coefficient for motion perpendicular to the axis (both drag coefficients have units of viscosity). In the limit that the filament becomes infinitely thin, we have $\zeta_{\perp} = 2 \zeta_{\parallel}$. The above representation is accurate to ${ O}(1/\log^{2}(R/L))$, where $R$ is the radius of the cylinder and $L$ is the total length of the cylinder \citep{Lauga2009}. The total hydrodynamic force, $\mathbf{F}$, and torque, $\mathbf{L}$, on the body from any motion can then be determined by integrating
\begin{eqnarray}
\mathbf{F} &=&\displaystyle\int_{-3}^{3} \mathbf{f} \,ds ,\\
L &=&\displaystyle\int_{-3}^{3} \mathbf{r}\wedge\mathbf{f} \,ds \cdot\mathbf{\hat{z}}',
\end{eqnarray}
where $\wedge$ is the wedge product and $\mathbf{\hat{z}}' = \mathbf{\hat{x}}' \wedge\mathbf{\hat{y}}' $ can be interpreted as the vector perpendicular to the plane of motion. This technique can be used to determine the force and torque on the body from Eq.~\eqref{def} and rigid body motion for arbitrary configurations. However these forces are very complex and so are not practical for demonstration. Hence we will only state the results for small $\phi_{1}$ and $\phi_{2}$ to keep the example tractable.

In the small-$\phi_{1}$ and $\phi_{2}$ limit the net force and torque from the deformation velocity, Eq.~\eqref{def}, is
\begin{eqnarray}
\mathbf{F}_{\phi} &=& \frac{4}{3} \Delta \zeta \left\{ (\phi_{2}-2 \phi_{1}) \dot{\phi_{1}} - (\phi_{1}-2 \phi_{2})\dot{\phi_{2}},-(\phi_{1}- \phi_{2})(\phi_{2} \dot{\phi_{1}}+\phi_{1} \dot{\phi_{2}})\right\} +{ O}(\phi^{4}),\\
L_{\phi} &=& -\frac{14}{3} \zeta_{\perp}(\dot{\phi_{1}}+\dot{\phi_{2}}) + \frac{1}{9}\left[4 (3 \bar{\zeta} + 5 \Delta \zeta)\phi_{1}^{2}-6 ( \bar{\zeta} + 5 \Delta \zeta)\phi_{1} \phi_{2}+( 3\bar{\zeta} + 19 \Delta \zeta)\phi_{2}^{2} \right]\dot{\phi_{1}} \notag \\
&& + \frac{1}{9}\left[(3 \bar{\zeta} + 19 \Delta \zeta)\phi_{1}^{2}-6 (\bar{\zeta} + 5 \Delta \zeta)\phi_{1} \phi_{2}+4( 3\bar{\zeta} + 5 \Delta \zeta)\phi_{2}^{2} \right]\dot{\phi_{2}} + { O}(\phi^{4}),
\end{eqnarray}
where $\bar{\zeta}= (\zeta_{\perp} +\zeta_{\parallel})/2$, and $\Delta \zeta = (\zeta_{\perp} -\zeta_{\parallel})/2$. 

Similarly the relationship between the force, $\mathbf{F}_{R} =\{F_{x},F_{y}\}$, and torque, $L_{R}$, and the rigid body linear velocity, $\mathbf{U}=\{U_{x},U_{y}\}$, and angular velocity $\Omega$ is given by
\begin{equation}
\left(\begin{array}{c}
F_{x} \\
F_{y} \\
L_{R}
\end{array}\right) = \left(\begin{array}{c c c}
A & B & C \\
B & D & E \\
C & E &F
\end{array} \right)\left(\begin{array}{c}
U_{x} \\
U_{y} \\
\Omega
\end{array}\right),
\end{equation}
where
\begin{eqnarray}
A &=& -6 \zeta_{\parallel}- 4 \Delta \zeta \left(\phi_{1}^{2}+\phi_{2}^{2} \right) + {\cal O}(\phi^{4}), \\
B &=& 4 \Delta \zeta \left(\phi_{1}+\phi_{2} \right) - \frac{8}{3}\Delta \zeta\left(\phi_{1}^{3}+\phi_{2}^{3} \right) + { O}(\phi^{4}), \\
C &=& 2 \Delta \zeta \left(\phi_{1}-\phi_{2} \right)\left(-4 +2\phi_{1}^{2}+\phi_{1}\phi_{2}+2\phi_{2}^{2} \right) + {\cal O}(\phi^{4}), \\
D &=& -6 \zeta_{\perp}+4 \Delta \zeta \left(\phi_{1}^{2}+\phi_{2}^{2} \right) +  {O}(\phi^{4}), \\
E &=& -\frac{16}{3}\Delta \zeta \left(\phi_{1}^{2}+\phi_{2}^{2} \right) + { O}(\phi^{4}), \\
F &=& -18 \zeta_{\perp} + \frac{4}{3}\left[2(\bar{\zeta}+4 \Delta \zeta)\phi_{1}^{2} - (\bar{\zeta}+7 \Delta \zeta)\phi_{1}\phi_{2}+2(\bar{\zeta}+4 \Delta \zeta)\phi_{1}^{2} \right] + {O}(\phi^{4}).
\end{eqnarray}

\subsubsection{The gauge field of Purcell's two-hinged swimmer }

The gauge field of the Purcell's two-hinged swimmer is determined through the balance of the forces and torques. If the swimmer is force and torque free, the instantaneous velocities are in the body frame
\begin{eqnarray}
U_{x} &=& -\frac{2 \Delta \zeta}{81 \zeta_{\parallel}}\left[(4 \phi_{1} + 5 \phi_{2})\dot{\phi_{1}}-(5 \phi_{1} + 4 \phi_{2})\dot{\phi_{2}}  \right]+ { O}(\phi^{3}), \\
U_{y} &=& 0 + { O}(\phi^{3}), \\
\Omega &=& -\frac{7}{27} (\dot{\phi_{1}}+\dot{\phi_{2}}) - \frac{1}{1458\zeta_{\parallel} \zeta_{\perp}} \left( \dot{\phi_{1}} Q (\phi_{1},\phi_{2}) + \dot{\phi_{2}} Q (\phi_{2},\phi_{1})\right) + { O}(\phi^{4}),
\end{eqnarray}
where
\begin{eqnarray}
Q(x,y) &=& (27 \Delta \zeta^{2}+ 24 \Delta \zeta \bar{\zeta} + 29 \bar{\zeta}^{2})y^{2} + 2 (13 \bar{\zeta}-15 \Delta \zeta)(\bar{\zeta}+3  \Delta \zeta)x y \notag \\
&& -4(27 \Delta \zeta^{2} - 24 \Delta \zeta \bar{\zeta} + 13 \bar{\zeta}^{2})x^{2}.
\end{eqnarray}
In the above, the angular velocity has been expanded to a higher order than the linear velocity  because the leading net rotation occurs at ${O}(\phi^{3})$ while the leading net translation occurs at ${ O}(\phi^{2})$. The body frame linear velocities are 0 if $\Delta \zeta=0$ while the rotation remains non-zero. This is consistent with the current understanding of the swimming of slender bodies \citep{Koens2016a}. 

The evolution of the laboratory frame position, $\mathbf{X}(t)= X(t) \mathbf{\hat{x}} + Y(t) \mathbf{\hat{y}}$, and orientation, $\theta(t)$, are found by expressing the velocities in the laboratory frame (Fig.~\ref{fig:purcell}a). In two dimensions the angular velocity remains unchanged but the translational body frame velocities are rotated. The rate of change of the configuration is 
\begin{eqnarray}
\left(\begin{array}{c}
\dot{X} \\
\dot{Y}\\
\dot{\theta}
\end{array} \right) &=&\left(\begin{array}{c c c}
\cos \theta & - \sin \theta & 0 \\
\sin \theta & \cos \theta & 0 \\
0&0&1 
\end{array} \right)\cdot\left(\begin{array}{c}
U_{x} \\
U_{y}\\
\Omega
\end{array} \right) \notag \\
&=& \mathbi{M}^{P} \cdot\left(\begin{array}{c}
\dot{\phi_{1}} \\
\dot{\phi_{2}}
\end{array} \right) , \label{labvelocity}
\end{eqnarray}
where
\begin{equation}
\mathbi{M}^{P} = \left(\begin{array}{c c}
  M_{x 1}^{P} & M_{x 2}^{P} \\
M_{y 1}^{P} & M_{y 2}^{P} \\
 M_{\theta 1}^{P} & M_{\theta 2}   ^{P}
\end{array} \right)
\end{equation}
and
\begin{eqnarray}
  M_{x 1}^{P} &=& -\frac{2 \Delta \zeta}{81 \zeta_{\parallel  } }(4 \phi_{1} + 5 \phi_{2}) \cos \theta,  \\
  M_{x 2}^{P} &=& \frac{2 \Delta \zeta}{81 \zeta_{\parallel  }}  (4 \phi_{2} + 5 \phi_{1}) \cos \theta,   \\
 M_{y 1}^{P} &=& -\frac{2 \Delta \zeta}{81 \zeta_{\parallel  }} (4 \phi_{1} + 5 \phi_{2}) \sin \theta,  \\
 M_{y 2}^{P} &=& \frac{2 \Delta \zeta}{81 \zeta_{\parallel  }} (4 \phi_{2} + 5 \phi_{1}) \sin \theta,   \\
 M_{\theta 1}^{P} &=& -\frac{7}{27} - \frac{Q (\phi_{1},\phi_{2})}{1458\zeta_{\parallel} \zeta_{\perp} }, \\
 M_{\theta 2}^{P} &=& -\frac{7}{27} - \frac{Q (\phi_{2},\phi_{1})}{1458\zeta_{\parallel} \zeta_{\perp} }. 
\end{eqnarray}
 Here $\mathbi{M}^{P}$ is the gauge field for the small-angled Purcell's two-hinged swimmer. The laboratory frame displacement can be determined by integrating Eq.~\eqref{labvelocity} for  prescribed $\phi_{1}$  and $\phi_{2}$. Swimming strokes have the additional constraint that $\phi_{1}$  and $\phi_{2}$ must be periodic, so the pattern can be repeated. The net displacements observed in the laboratory frame are therefore given by
\begin{equation}
\left(\begin{array}{c}
\Delta X \\
\Delta Y\\
\Delta \theta
\end{array} \right)  = \displaystyle\oint \mathbi{M}^{P} \cdot \frac{d\boldsymbol{l}_{P}}{dt}  \,dt  = \displaystyle\oint \mathbi{M}^{P} \cdot \,d\boldsymbol{l}_{P} ,
\end{equation}
 where $\boldsymbol{l}_{P} = \{\phi_{1},\phi_{2}\} $ are the two deformation modes of the Purcell swimmer. This result is independent of the speed at which $\boldsymbol{l}_{P}$ varies, as expected for swimmers in Stokes flow \citep{Purcell}. 

\subsubsection{Rotation of small-angled Purcell's two-hinged swimmer}

The net rotation of Purcell's two-hinged swimmer can be determined with the extended geometric swimmer techniques since the rotation part of the gauge field ${\mathbf{M}^{P}_{\theta}=\{ M_{\theta 1}^{P},  M_{\theta 2}^{P}\}}$ is independent of the swimmers position and orientation. When separated from the translations, the net angular displacement for the Purcell swimmer, $\Delta \theta$, is expressed as
\begin{equation}
\Delta \theta = \displaystyle\oint \mathbf{M}_{\theta}^{P}(\mathbf{l}_{P})\cdot\,d\boldsymbol{l}_{P}.
\end{equation}
This equation is a path integral and can be determined if we prescribe a specific stroke. For example, the Purcell-like stoke
\begin{equation}
\boldsymbol{\phi}_{ex} = \left\{ \begin{array}{c r}
\{ a-A + 2 A t, -A\} &   0<t<1, \\
\{a + A, -A + 2 A (t-1)\} &   1<t<2, \\
\{ a+A -2 A (t-2), A\} &   2<t<3, \\
\{a - A, A - 2 A (t-3)\} &   3<t<4,\\
\end{array} \right. \label{exloop1}
\end{equation} 
where $a$ and $A$ are constants, generates a net rotation of
\begin{equation}
\Delta \theta_{ex} =  \displaystyle\oint \mathbf{M}_{\theta}^{P}(\boldsymbol{\phi}_{ex})\cdot\,d\boldsymbol{\phi}_{ex} =  \frac{32 a A^{2} (2 \bar{\zeta}^{2} + 9\Delta\zeta^{2}) }{729 \zeta_{\parallel} \zeta_{\perp} }.
\end{equation}
The loop $\boldsymbol{\phi}_{ex}$ forms a square with side lengths $2A$ in phase space offset from the origin by $a$ in $\phi_1$ and becomes the Purcell swimming stroke when $a=0$ \citep{Purcell, BECKER2003}. In the $a=0$ limit, the rotation goes to zero because of the symmetries of the motion. 

Multiple evaluations of these path integrals are needed to determine the behaviour of the swimmer. This can be quite cumbersome and so it is favourable to use Green's theorem to transform the net rotation into
\begin{eqnarray}
\Delta \theta &=& \displaystyle\oint_{\partial V} \mathbf{M}_{\theta}^{P}(\mathbf{l}_{P})\cdot\,d\boldsymbol{l}_{P} \notag \\
&=& \iint_{V} \left(\frac{\partial M_{\theta 2}}{\partial \phi_{1}} - \frac{\partial M_{\theta 1}}{\partial \phi_{2}}\right) \,d \phi_{1} \,d\phi_{2}  \notag \\
&=&\frac{32  (2 \bar{\zeta}^{2} + 9\Delta\zeta^{2}) }{729 \zeta_{\parallel} \zeta_{\perp} } \iint_{V} \left( \phi_{1}-\phi_{2}\right)\,d \phi_{1} \,d\phi_{2}, \label{rotG}
\end{eqnarray} 
where the above double integrals are to be taken over the area of the loop.  This integral is much simpler to deal with than the field $\mathbf{M}_{\theta}^{P}(\mathbf{l}_{P})$ because the components in $\mathbf{M}_{\theta}^{P}(\mathbf{l}_{P})$ that do not contribute to the net displacement are removed through taking   derivatives. This makes the evaluation the integral significantly easier. 
For example the net rotation from loop $\boldsymbol{\phi}_{ex}$ is
\begin{equation}
\Delta \theta_{ex} =\frac{32  (2 \bar{\zeta}^{2} + 9\Delta\zeta^{2}) }{729 \zeta_{\parallel} \zeta_{\perp} } \int^{a+A}_{a-A}  \int_{-A}^{A} \left( \phi_{1}-\phi_{2}\right)\,d\phi_{2}\,d\phi_{1}  = \frac{32 a A^{2} (2 \bar{\zeta}^{2} + 9\Delta\zeta^{2}) }{729 \zeta_{\parallel} \zeta_{\perp} },
\end{equation}
identically to the path integral evaluation. 

This relationship holds for any loop and so the net rotation can be estimated through a plot of the integrand (see Fig.~\ref{fig:purcell}b). The ability to quickly estimate the net rotation by eye provides a practical method to design strokes with desired rotations. Since regions that can generate larger rotations can be identified simply by evaluating the magnitude of the plot. 

Though this technique has worked well to determine the net rotation, the translational terms depend on the orientation of the swimmer in the laboratory frame, $\theta$. This provides the integral with a memory of the path taken and thus prevents the use of Green's theorem directly. { As a result many have used the series approximation to treat such problems \citep{Hatton2011, Hatton2013, Hatton2015, Ramasamy2017, Hatton2017, Ramasamy2019, Bittner2018}. Alternatively, we show that this problem can be overcome exactly by embedding the integral into a suitable higher dimensional space. This is needed to generally discuss the behaviour of any swimmer's displacement.}

\section{Treatment of non-commuting variables} \label{sec:commute}

The generalized Stokes theorem can be used whenever the displacement is given by
\begin{equation}
\Delta \mathbf{x} = \oint_{\partial V} \mathbi{M}(\mathbf{l})\cdot d \mathbf{l},
\end{equation}
where $\mathbi{M}(\mathbf{l})$ only depends on the configuration of the swimmer. However in general the field $\mathbi{M}(\mathbf{l},\mathbf{x})$ depends on both the swimmers position and orientation through $\mathbf{x}$. This dependence can arise from many sources, such as the influence of boundaries in the swimmer's resistance matrix, Eq.~\eqref{RM}, the fact that rotations and translations do not commute, or hydrodynamics interactions between multiple swimmers. This additional dependence on $\mathbf{x}(t)$ gives the field a memory of the path taken and so the Stokes theorem no longer applies. As discussed above, corrections to the Stokes theorem can be formed but typically restrict the solution to smaller loops. In this section, we describe a general method to overcome the presence of non-commuting variables exactly. This is achieved by embedding the system in a higher dimensional space in which these variables are treated as prescribed paths and the generalized Stokes theorem can be used again after closing the paths. This embedding method has been used previously to model the swimming from a non-neutrally buoyant scallop \citep{Burton2010}. We generalise this idea to any swimmer and any stroke. This physically allows for strokes in which the swimmer experiences external forces and torques and thereby enables the embedded representation to treat the dynamics of a swimmer in any scenario simultaneously. Any specific case exists as a subspace of the whole embedded space. We first derive this representation in its most general form and then demonstrate its equivalence to path integral representation using the translation of the Purcell swimmer

\subsection{General embedding of non-commuting variables}

 \begin{figure}
\centering
\includegraphics[width=\textwidth]{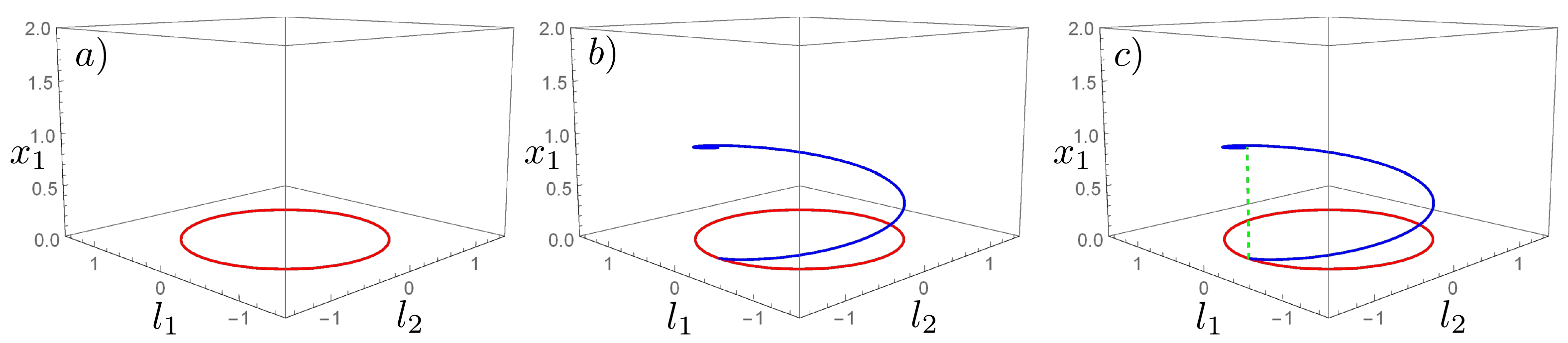}
\caption{Diagram depicting the embedding of displacement integral into higher space. a) The original path (red) used for the integral in the configuration plane. b) The equivalent path (dark blue) after treating the non-commuting variables as a specified path. c) The dashed path (green) used to close the loop.  } 
\label{fig:emb}
\end{figure} 

Consider a general swimmer whose position and orientation, $\mathbf{x}$, satisfy Eq.~\eqref{evo}. The position of the swimmer at each point in time can be determined by integrating this equation with respect to a prescribed path, Eq.~\eqref{pos} and the net displacement is found when this path forms a closed loop, Eq.~\eqref{net} (Fig.~\ref{fig:emb}a). In a parametric sense this net displacement can also be written as
\begin{equation}
\Delta\mathbf{x} = \int_{0}^{t_{fin}} \mathbi{M}(\mathbf{l}(t), \mathbf{x}(t))\cdot \frac{d \mathbf{l}}{dt} \,dt, \label{path0}
\end{equation}
where the loop is parametrised such that $\mathbf{l}(0)=\mathbf{l}(t_{fin})$ and $t \in [0,t_{fin}]$. Importantly, $\Delta\mathbf{x}$ is independent of the choice of parametrisation as required by the time-independence of the Stokes equations. This integral appears to depend on two parametrised paths $\mathbf{l}(t)$ and $\mathbf{x}(t)$. This consideration might prompt us to rewrite the integral formally as
\begin{eqnarray}
\Delta\mathbf{x} &=& \int_{0}^{t_{fin}} \left(\mathbi{M}(\mathbf{l}(t), \mathbf{x}(t))\cdot \frac{d \mathbf{l}}{dt} + \mathbf{0}\cdot \frac{d \mathbf{x}}{dt}  \right) \,dt \notag \\
&=& \int_{0}^{t_{fin}}  \mathbi{M}'(\mathbf{l}'(t))\cdot \frac{d \mathbf{l}'}{dt}\,dt \label{presc}
\end{eqnarray}
where $\mathbf{l}' = \{\mathbf{l},\mathbf{x}\}$ and $\mathbi{M}' = \{\mathbi{M},\mathbi{0}\}$. This reformulation treats $\mathbf{x}(t)$ as a prescribed path by embedding the path integral into a higher dimensional space (see Fig.~\ref{fig:emb}b). The dimension of the new configuration space is the number of deformation modes, $\mathbf{l}$, plus the number of position and orientation variables, $\mathbf{x}$. The prescription of $\mathbf{x}(t)$ removes any path memory and therefore makes the representation closer to what is required to apply the generalized Stokes theorem. However it also allows for paths that do not satisfy Eq.~\eqref{evo}. These new paths correspond to systems in which the swimmer is subject to external forces and torques \citep{Burton2010}. Biological microscopic swimmers often experience such external forces and torques through gravity or the additional drag produced by background fluid flows \citep{Martinez2020, Bianchi2017, Goldstein2015, PerezIpina2019, Gaffney2011, Lauga2016} while several artificial microscopic swimmers rely on external forces and torque to drive the motion  \citep{Huang2019, Zhang2010, Vizsnyiczai2017}. Hence these these additional paths allows us to consider the behaviour of the swimmer in any environment simultaneously and will enable us to show that the displacement of the swimmer in any scenario can always be visualised on a single surface. The paths which correspond to a specific physical scenario form a subset of this space. This subset is model-dependant and simply connected if variations of the path produce a continuous variation of the displacement. The general identification of these regions has been left for future work, however, we note that in some situations these paths can be found by solving some of the governing equations \citep{Burton2010}.

The generalized Stokes theorem requires a closed path integral over a field that only depends on the prescribed variables at that point of space. The above embedding has removed the memory contributions from the non-commuting variables $\mathbf{x}(t)$. However $\mathbf{x}(t)$ is expected to change over the course of a stroke. This means that the path in Eq.~\eqref{presc} is not closed (Fig.~\ref{fig:emb}b). The difference between the start and end position in this higher dimensional space is
\begin{equation}
\mathbf{l}'(t_{fin}) - \mathbf{l}'(0) = \{ \mathbf{l}(t_{fin})-\mathbf{l}(0),\mathbf{x}(t_{fin})-\mathbf{x}(0)\} = \{\mathbf{0},\mathbf{x}(t_{fin})-\mathbf{x}(0)\},
\end{equation}
  where we have used the periodicity condition for $\mathbf{l}(t)$. The difference between the start and ends of the new path occurs only in the added dimensions and path movements in these added dimensions contributes nothing to the final integral. Therefore if the original path is given by $\mathbf{l}'(t)= \{\mathbf{l}(t),\mathbf{x}(t)\}$, the path integral of $\mathbf{k}'(t)=\{\mathbf{l}(0), \mathbf{x}(t_{fin}-t)\}$ gives
  \begin{equation}
  \int_{0}^{t_{fin}}  \mathbi{M}'(\mathbf{k}'(t))\cdot \frac{d \mathbf{k}'}{dt}\,dt = \int_{0}^{t_{fin}} \left(\mathbi{M}(\mathbf{l}(t), \mathbf{x}(t))\cdot \mathbf{0} + \mathbi{0}\cdot \frac{d \mathbf{x}}{dt}  \right) \,dt ={\bf 0},
  \end{equation}
and has a distance between the start and end of $\mathbf{k}'(t_{fin})-\mathbf{k}'(0) =   \{\mathbf{0},-\mathbf{x}(t_{fin})+\mathbf{x}(0)\}$. This $\mathbf{k}'$ path starts at $\mathbf{l}'(t_{fin})$, ends at $\mathbf{l}'(0)$ and does not contribute to the total displacement. Hence the net displacement integral can then be written as
\begin{eqnarray}
\Delta\mathbf{x}  &=& \int_{0}^{t_{fin}}  \mathbi{M}'(\mathbf{l}'(t))\cdot \frac{d \mathbf{l}'}{dt}\,dt + \int_{0}^{t_{fin}}  \mathbi{M}'(\mathbf{k}'(t))\cdot \frac{d \mathbf{k}'}{dt}\,dt \notag \\
&\equiv& \oint_{\mathbf{l}' + \mathbf{k}'} \mathbi{M}'(\mathbf{l}')\cdot \,d\mathbf{l}',
\end{eqnarray}
  which again closes the integral (Fig.~\ref{fig:emb}c). We emphasize that the path $\mathbf{k}'(t)$ was chosen to travel along ${\mathbf{x}(t_{fin}-t)}$ intentionally. In principle several paths can be used to close this loop in the higher dimensional space. However in periodic geometries these closing paths could generate a contribution from the periodicity. This is avoided by making $\mathbf{k}'(t)$ travel backwards along $\mathbf{x}(t)$. Crucially, the above loop closure has turned the net displacement integral into a form in which the generalized Stokes theorem can be directly applied. { This embedding process is simple enough to be done by hand and overcomes the influence of non-commuting variables exactly without the need to assume small swimming strokes. }
  
 \begin{figure}
\centering
\includegraphics[width=0.5\textwidth]{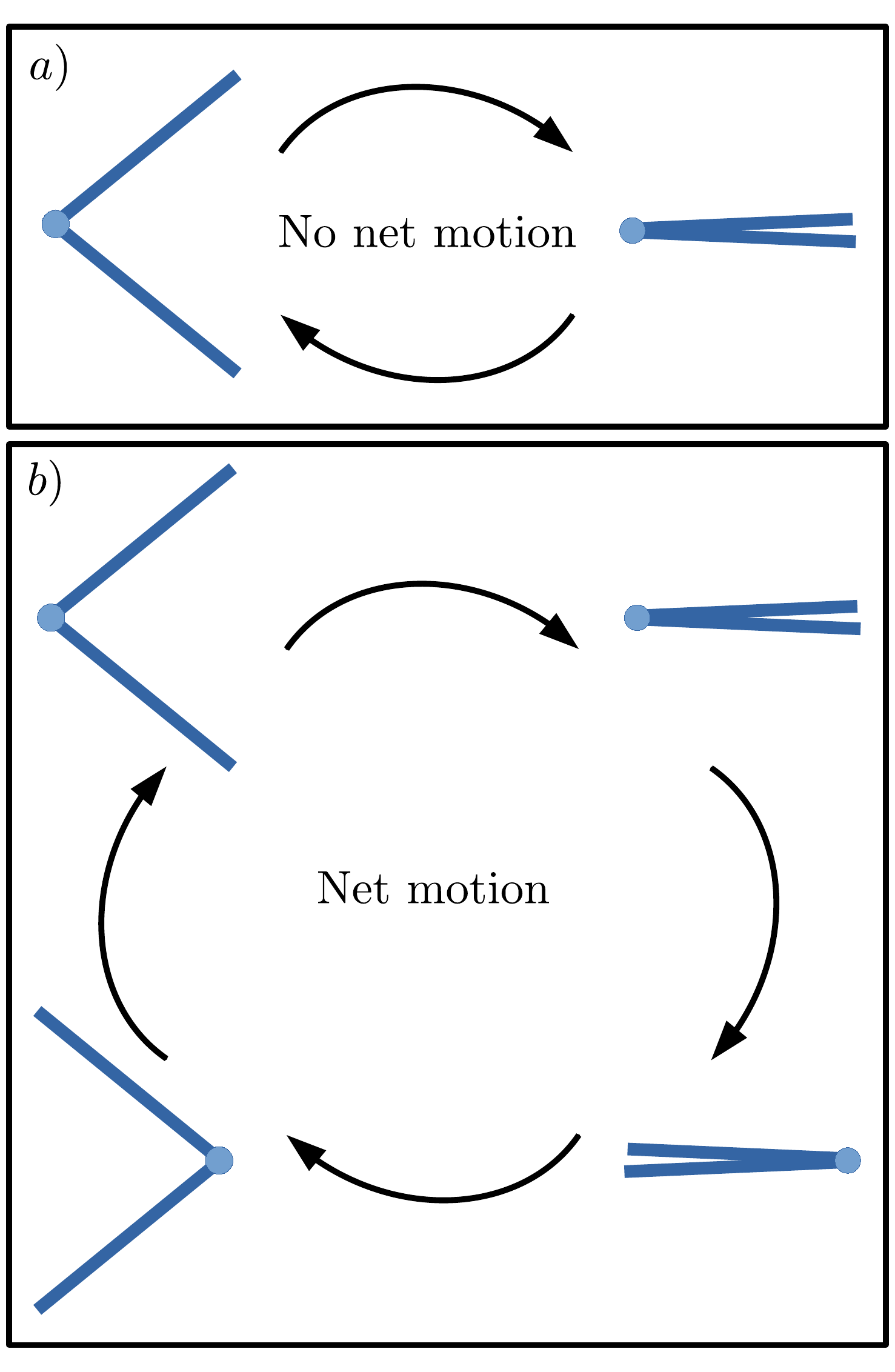}
\caption{Diagrams depicting how a one-hinged swimmer can move. a) A simple open and close motion produces no net motion due to the scallop theorem. b) The same open and close motion combined with laboratory frame rotations can allow net motions. } 
\label{fig:scallop}
\end{figure}

The ability to treat $\mathbf{x}$ and $\mathbf{l}$ in this way, and thereby invoke the generalized Stokes theorem, is a reflection of how these factors interact to generate motion. Consider a reciprocal stroke in $\mathbf{l}$. Without a change in $\mathbf{x}$ the stroke will generate no translation. However if $\mathbf{x}$ changes suitably throughout the deformation, displacement can occur. This has been demonstrated by \citet{Burton2010} who considered a single hinged swimmer (scallop) which could change its centre of buoyancy throughout the stroke. Though the swimmer cannot swim by opening an closing the hinge under normal circumstances, by changing its centre of buoyancy the swimmer rotated itself throughout the stroke and so generated a net displacement (Fig.~\ref{fig:scallop}). The full range of net displacements possible requires therefore the consideration of all combinations of $\mathbf{x}$ and $\mathbf{l}$. 

 The above derivation proves that the path integral for the displacement, Eq.~\eqref{path0}, is equivalent to a closed path integral representation in a $\mathbf{l}+\mathbf{x}$ dimensional space without path memory terms. Hence the displacement of any swimmer, in any environment, can always be written in a form where the generalized Stokes theorem can be applied. The ability to always write the displacement in this way has specific implications about the motion of viscous microswimmers and will allow all possible displacements of a swimmer to be identified through a single surface, as we will show in Sec.~\ref{sec:vis}.

\subsection{Example: Translation of a Purcell swimmer}

\subsubsection{Small-hinge angle demonstration}
The  equivalence of the above representations can be demonstrated { analytically} by considering the net translation of the Purcell swimmer in the small-angle limit. The translation of the swimmer, in this limit, is given by
\begin{equation}
\left(\begin{array}{c}
\Delta X \\
\Delta Y\\
\end{array} \right)  = \displaystyle\oint \left(\begin{array}{c c}
M_{x1}^{P}(\phi_{1},\phi_{2},\theta) & M_{x2}^{P}(\phi_{1},\phi_{2},\theta) \\
 M_{y1}^{P}(\phi_{1},\phi_{2},\theta) & M_{y2}^{P}(\phi_{1},\phi_{2},\theta)
\end{array} \right) \cdot 
\displaystyle \left(\begin{array}{c}
\displaystyle \frac{d \phi_{1}}{dt}  \\
\displaystyle  \frac{d \phi_{2}}{dt} 
\end{array}\right)  \,dt. \label{trans}
\end{equation}
 In the absence of external forces and torques, trajectories for the orientation, $\theta$, satisfy Eq.~\eqref{labvelocity} { which to the accuracy we have considered for translation, $ O(\phi^{2})$, has the solution}
\begin{equation}
\theta(t) = \theta_{0} - \frac{7[\phi_{1}(t)+\phi_{2}(t)]}{27} +\frac{7[\phi_{1}(0)+\phi_{2}(0)]}{27} + O(\phi^{3}), \label{leading theta}
\end{equation}  
where $\theta_{0}$ is the initial orientation of the swimmer. { The restriction on the accuracy of $\theta$ is necessary to ensure we do not expand the model inconsistently when considering translation. The above equation shows that to this order $\theta- \theta_0$ is $O(\phi)$ and so must also be treated as a small parameter in the small angle limit.} In which case the relevant components of the $\mathbi{M}$ become
\begin{eqnarray}
  M_{x 1}^{P} &=& -\frac{2 \Delta \zeta}{81 \zeta_{\parallel  } }(4 \phi_{1} + 5 \phi_{2}) \left[\cos \theta_{0} - (\theta-\theta_{0}) \sin \theta_{0}\right],  \\
  M_{x 2}^{P} &=& \frac{2 \Delta \zeta}{81 \zeta_{\parallel  }}  (4 \phi_{2} + 5 \phi_{1}) \left[\cos \theta_{0} - (\theta-\theta_{0}) \sin \theta_{0}\right],   \\
 M_{y 1}^{P} &=& -\frac{2 \Delta \zeta}{81 \zeta_{\parallel  }} (4 \phi_{1} + 5 \phi_{2}) \left[\sin \theta_{0} + (\theta-\theta_{0}) \cos \theta_{0}\right],  \\
 M_{y 2}^{P} &=& \frac{2 \Delta \zeta}{81 \zeta_{\parallel  }} (4 \phi_{2} + 5 \phi_{1}) \left[\sin \theta_{0} + (\theta-\theta_{0}) \cos \theta_{0}\right].
\end{eqnarray}
The net displacement generated from the Purcell like loop, $\boldsymbol{\phi}_{ex}$, (Eq.~\ref{exloop1}) in this limit is
\begin{equation}
\left(\begin{array}{c}
\Delta X \\
\Delta Y\\
\end{array} \right)  = - \frac{8  A^{2} \Delta \zeta }{2187 \zeta_{\parallel}}\left(\begin{array}{c}
 270 \cos \theta_{0} +7 (9a + 20 A) \sin \theta_{0} \\
 270 \sin \theta_{0} -7  (9a + 20 A)  \cos \theta_{0} \\
\end{array} \right). \label{example1}
\end{equation}
The non-zero $\Delta Y$ for $\theta_{0}=0$ arises solely from the interactions of rotations and translations.  When $a=\theta_0=0$, this loop corresponds to the Purcell swimming stroke \citep{Purcell} and the displacement is identical to the small angled asymptotic results found by \citet{BECKER2003}.

The dependence on the orientation of the swimmer, $\theta$, of the net displacements means that Green's theorem cannot be applied to the integral directly. As discussed above this can be overcome by treating $\theta$ as a prescribed path, and reformating the integrals as
\begin{eqnarray}
\left(\begin{array}{c}
\Delta X \\
\Delta Y\\
\end{array} \right)  &=& \displaystyle\oint_{\partial V} \left(\begin{array}{c c c}
M_{x1}^{P}(\phi_{1},\phi_{2},\theta) & M_{x2}^{P}(\phi_{1},\phi_{2},\theta) & 0 \\
 M_{y1}^{P}(\phi_{1},\phi_{2},\theta) & M_{y2}^{P}(\phi_{1},\phi_{2},\theta) & 0
\end{array} \right) \cdot \left(\begin{array}{c}
\displaystyle \frac{d \phi_{1}}{dt}  \\
\displaystyle \frac{d \phi_{2}}{dt}  \\
\displaystyle \frac{d \theta}{dt}
\end{array}\right)  \,dt, \notag \\
&=& \displaystyle\oint_{\partial V} \left(\begin{array}{c }
\mathbf{M'}^{P}_{x}(\phi_{1},\phi_{2},\theta)\\
\mathbf{M'}^{P}_{y}(\phi_{1},\phi_{2},\theta)
\end{array} \right) \cdot d \mathbf{l}_{P}'
\end{eqnarray}
where $\partial V$ is the loop considered, $\mathbf{M'}^{P}_{x} = \{ M_{x1}^{P}(\phi_{1},\phi_{2},\theta), M_{x2}^{P}(\phi_{1},\phi_{2},\theta) , 0\}$, \quad $\mathbf{M'}^{P}_{y} = \{ M_{y1}^{P}(\phi_{1},\phi_{2},\theta), M_{y2}^{P}(\phi_{1},\phi_{2},\theta) , 0\}$ and $d \mathbf{l}_{P}' =\{d \phi_{1}, d \phi_{2}, d \theta\}$. This form allows us to use Stokes theorem and so the net displacement can be written as
\begin{equation}
\left(\begin{array}{c}
\Delta X \\
\Delta Y\\
\end{array} \right)= \displaystyle\iint_{V} \left(\begin{array}{c }
\nabla \times \mathbf{M'}^{P}_{x}(\phi_{1},\phi_{2},\theta)\\
\nabla \times  \mathbf{M'}^{P}_{y}(\phi_{1},\phi_{2},\theta)
\end{array} \right) \cdot d \mathbf{S'}^{P}
\end{equation}
where $V$ is any surface bounded by $\partial V$,  $d \mathbf{S}_{P}'$ is the infinitesimal surface element,  $\nabla \times$ is the curl operation taken with respect to the coordinates $\{\phi_{1},\phi_{2},\theta\}$ and
\begin{eqnarray}
\nabla \times \mathbf{M'}^{P}_{x}(\phi_{1},\phi_{2},\theta) &=& \frac{2 \Delta \zeta}{81 \zeta_{\parallel}} \{(5 \phi_{1} + 4 \phi_{2}) \sin \theta_{0}  , (4 \phi_{1} + 5 \phi_{2})\sin \theta_{0}  , \notag  \\ 
&& \qquad \mbox{ }\quad 10 [\cos \theta_{0} - (\theta-\theta_{0}) \sin \theta_{0}]  \}, \label{pxdis}   \\
\nabla \times  \mathbf{M'}^{P}_{y}(\phi_{1},\phi_{2},\theta) &=& -\frac{2 \Delta \zeta}{81 \zeta_{\parallel}} \{ (5 \phi_{1} + 4 \phi_{2}) \cos \theta_{0} , (4 \phi_{1} + 5 \phi_{2})\cos \theta_{0} , \notag \\
&& \qquad \mbox{ }\quad -10 [ (\theta-\theta_{0}) \cos \theta_{0}- \sin \theta_{0}]\}. \label{pydis}  
\end{eqnarray}
 The $\nabla \times \mathbf{M'}^{P}_{x}(\phi_{1},\phi_{2},\theta)$ and $\nabla \times \mathbf{M'}^{P}_{y}(\phi_{1},\phi_{2},\theta)$ fields have been plotted in Fig.~\ref{fig:curlfields}.

 \begin{figure}
\centering
\includegraphics[width=0.9\textwidth]{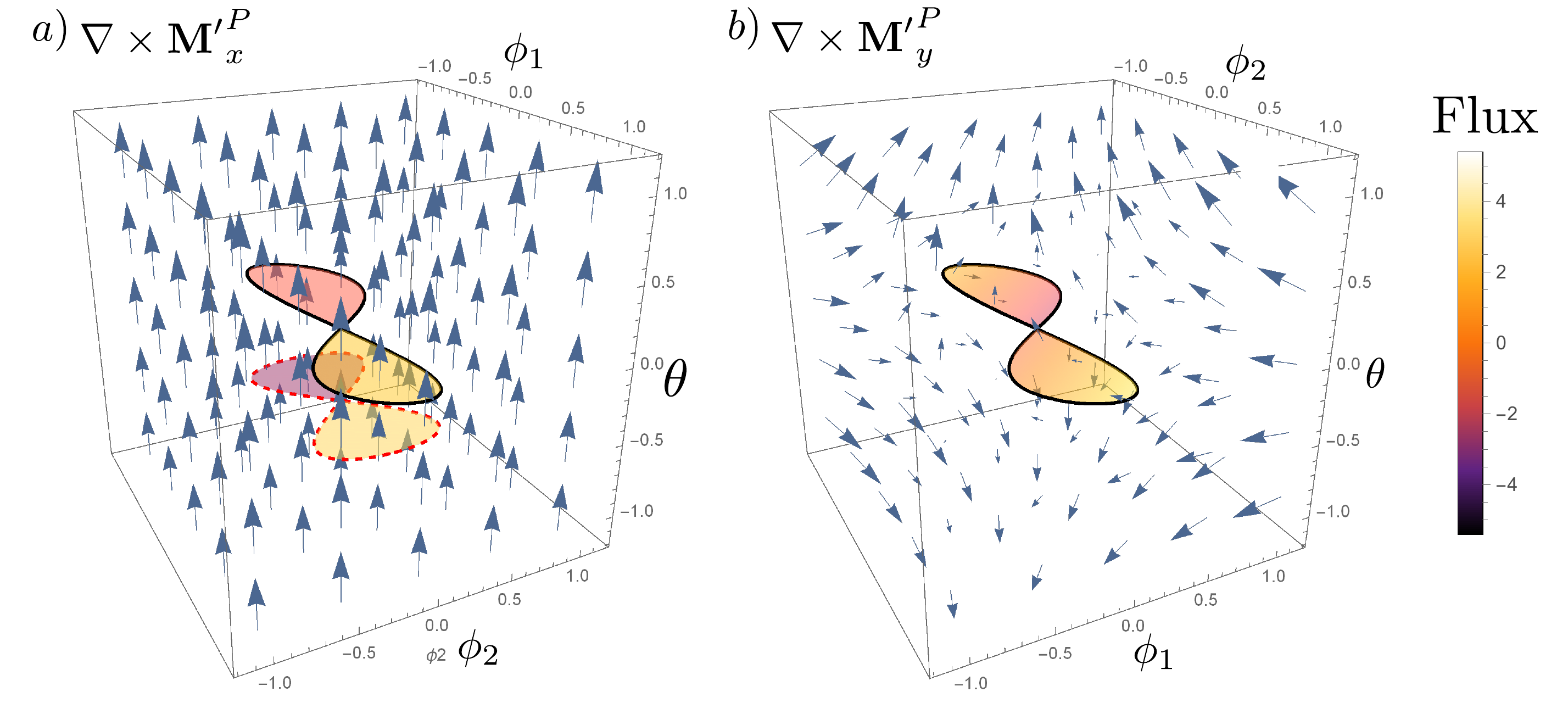}
\caption{The vector fields found from the curl of the displacement fields when $\theta_{0}=0$ for Purcell's swimmer in the small angle limit. a)$\nabla \times\mathbf{M'}^{P}_{x}(\phi_{1},\phi_{2},\theta)$ b) $\nabla \times \mathbf{M'}^{P}_{y}(\phi_{1},\phi_{2},\theta)$. The size of the arrows display the relative strength of the field at that point.  In both plots the stroke corresponding to $\phi_1 = 0.7 \cos(t)$, $\phi_2 =0.4 \sin(2 t)$ has been plotted in black with a surface that is bounded by said loop. The colour of this surface reflects the flux of field through the surface at each point with the total displacement being the total flux through the surface. The red dashed loop in a) generates the same $\Delta X$ as the black loop but exists solely in a $\phi_1$-$\phi_2$ plane.} 
\label{fig:curlfields}
\end{figure} 

The equivalence of this embedded representation and the path integral can be demonstrated with the example loop $\boldsymbol{\phi}_{ex}$. In this new space the example loop becomes $\boldsymbol{\phi}_{ex}' = \{\boldsymbol{\phi}_{ex},\theta(t)\}$, where we must use Eq.~\eqref{leading theta} for $\theta(t)$. This path is periodic and so does not need to be closed with another path. A surface that bounds this loop can be parametrised using $\phi_1$ and $\phi_2$ by $\mathbf{S'}^{P}_{ex}(s,t) = \{\phi_1, \phi_2,\theta_{0} - 7 (\phi_1 + \phi_2 -a + 2 A)/27 \}$ and so the surface element is  
\begin{equation}
d \mathbf{S'}^{P}_{ex} = \frac{\partial \mathbf{S'}^{P}_{ex}}{\partial \phi_1} \times \frac{\partial \mathbf{S'}^{P}_{ex}}{\partial \phi_2} \,d\phi_1 \,d\phi_2
= \left\{ \frac{7}{27} ,  \frac{7}{27},1\right\}\,d\phi_1 \,d\phi_2
\end{equation}
 The displacement is therefore given by 
\begin{eqnarray}
\left(\begin{array}{c}
\Delta X \\
\Delta Y\\
\end{array} \right)&=& \displaystyle \int_{-A}^{A} \,d\phi_2\int_{a-A}^{a+A} \,d\phi_1  \left(\begin{array}{c }
\nabla \times \mathbf{M'}^{P}_{x}(\phi_{1},\phi_{2},\theta)\\
\nabla \times  \mathbf{M'}^{P}_{y}(\phi_{1},\phi_{2},\theta)
\end{array} \right) \cdot  \left\{ \frac{7}{27} ,  \frac{7}{27},1\right\}\notag \\
  &=&- \frac{8  A^{2} \Delta \zeta }{2187 \zeta_{\parallel}}\left(\begin{array}{c}
 270 \cos \theta_{0} +7 (9a + 20 A) \sin \theta_{0} \\
 270 \sin \theta_{0} -7  (9a + 20 A)  \cos \theta_{0} \\
\end{array} \right),
\end{eqnarray}
which is identical to the result in Eq.~\eqref{example1}.  This result is independent of the surface chosen. However this freedom of choice can make the surface integral evaluation more complicated than the path integral for prescribed paths. We note that if $\boldsymbol{\phi}_{ex}'$ was not periodic, an additional path, such as that described above, can always be added to the path integral by tracing back on the $\theta(t)$ trajectory while leaving the other coordinates unchanged. Hence this representation allows Stokes theorem to be used for any stroke.

 The visualisation of the $\nabla \times \mathbf{M'}^{P}_{x}(\phi_{1},\phi_{2},\theta)$ and $\nabla \times \mathbf{M'}^{P}_{y}(\phi_{1},\phi_{2},\theta)$ fields (Fig.~\ref{fig:curlfields}) can be useful in the design of strokes for specific motions. Regions capable of producing higher displacement can be identified by the respective size of the arrows and loops can be aligned to control the flux of these arrows through them. For example, imagine we want to find strokes which produces $\Delta X =0$ but allows $\Delta Y \neq 0$ with $\theta_{0}=0$. We could prescribe a general $\phi_1$ and $\phi_2$ and try to find see when the path integrals evaluate to zero. However this is not necessary if we inspect the fields.  Since $\nabla \times \mathbf{M'}^{P}_{x}(\phi_{1},\phi_{2},\theta)$ is constant and solely in the $\theta$ direction, the value of $\Delta X$ from any stroke is related to the area of the loop projected on a $\phi_1$-$\phi_2$ plane. Regions of this projected loop traced counter-clockwise add to the displacement while regions traced clockwise subtract from it. Any combination of loops that trace the same area clockwise as counter-clockwise therefore produces no displacement. A figure-8 loop is a simple example of such a shape (Fig.~\ref{fig:curlfields}). These loops can however produce $\Delta Y \neq 0$ because $\nabla \times \mathbf{M'}^{P}_{x}(\phi_{1},\phi_{2},\theta)$ has a different structure and so meets our original criteria. Hence Fig.~\ref{fig:curlfields} has allowed us to quickly identify a large set of strokes that meet our criteria without the need of any calculations and so can be useful for designing strokes. Unfortunately the visualisation of such fields in dimensions higher than three is tricky and so different methods need to be produced. In the next section we produce a method to visualise the displacements from any loop in the extended space on a surface, similarly to how $\Delta X$ could be determined by considering the area enclosed by the projected loop on a $\phi_1$-$\phi_2$ plane.

{ 
\subsubsection{Comparison with the minimal perturbation coordinates series representation}

The small-hinge angle translation model used for demonstration, is equivalent to the small-stroke size limit of the Lie bracket approximation, Eq.~\eqref{Lie}. As a result, the net displacement from the leading approximation in the minimal perturbation coordinates will be the same as the embedded representation. A comparison of the accuracy between the embedded method and the small-stroke approximation in the minimal perturbation coordinates \citep{Hatton2011, Hatton2013, Hatton2015, Ramasamy2017, Hatton2017, Ramasamy2019, Bittner2018} requires us therefore to consider the Purcell swimmer problem with arbitrary sized hinge angles. This full system can be done analytically for the embedded representation but is rather complex and so we have omitted it here for brevity. A version of the numerical program necessary for the small-stroke approximation in the minimal perturbation coordinates is freely available and works in MATLAB \citep{hatton_2020}. We compared the results from this program to the exact results for the square Purcell strokes to ensure we ran it correctly.  Note that in the full Purcell swimmer system the $y$ displacement field is the same as the $x$ field but with a phase shift of $\theta = \pi/2$.

We compared the displacement found from these methods for strokes of the form 
\begin{equation}
\{\phi_1,\phi_2\} = \{\Phi_1 + \cos(t), \Phi_2 + \sin(t)\}, \label{comp}
\end{equation}
for varying $\Phi_1 $ and $\Phi_2$. These strokes are unit circles in the configuration space centred about $\{\Phi_1,\Phi_2\}$.  Contour plots of the net displacement in $x$ and $y$ determined by solving Eq.~\eqref{evo}, the minimal perturbation coordinate approximation, the embedded results using the exact form of $\theta(t)$ and the embedded results for a guess $\theta_g(t) = t \Delta \theta / 2 \pi$ are shown in Fig.~\ref{fig:compare}. The guess $\theta_g(t)$ is the simplest from which accounts for a change in angle over the stroke. Unsurprisingly the embedded results with the exact form of $\theta(t)$ (Fig.~\ref{fig:compare}$e$,$f$) is identical to the solution from Eq.~\eqref{evo} (Fig.~\ref{fig:compare}$a$,$b$) as we proved that it is mathematically equivalent above. The minimal perturbation coordinate approximation (Fig.~\ref{fig:compare}$c$,$d$) and the embedded results with $\theta_g(t)$  (Fig.~\ref{fig:compare}$g$,$h$) both replicate the $x$ displacement better than the $y$ displacement but in each case have regions with over 10\% error and sometimes get the sign of the displacement wrong. Furthermore the similarity in the structure of the embedded results with $\theta_g(t)$ (Fig.~\ref{fig:compare}$g$,$h$) and the solution from Eq.~\eqref{evo} (Fig.~\ref{fig:compare}$a$,$b$) could be a coincidental result specific to the Purcell swimmer. {\color{blue} We note that the exponentiated map of the minimal perturbation coordinate approximation has a negligible difference with the exact solution for the values tested.}

 \begin{figure}
\centering
\includegraphics[width=0.95\textwidth]{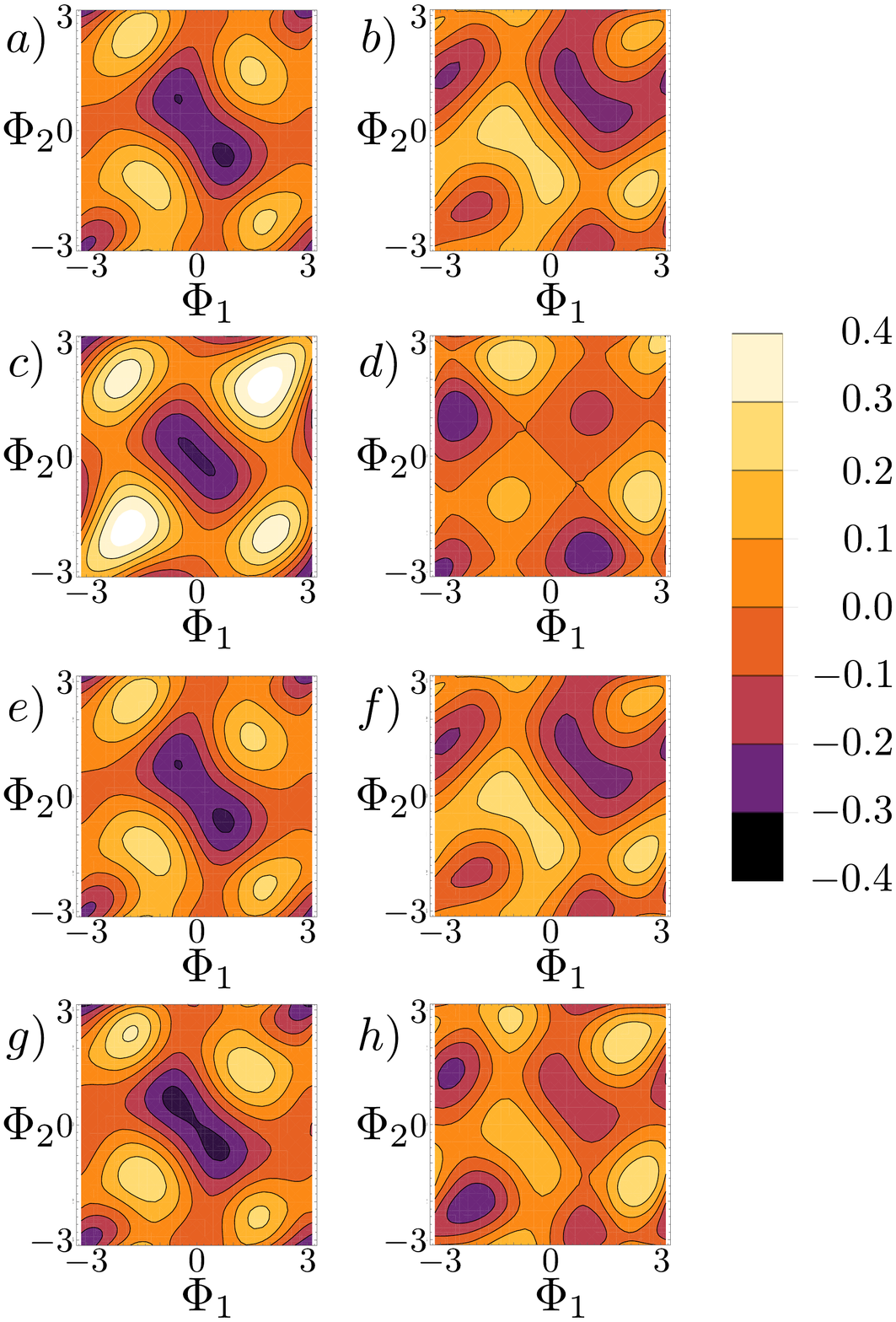}
\caption{{Density plots of the net displacement in $x$ ($a$, $c$, $e$, $g$) and $y$ ($b$, $d$, $f$, $h$) from a Purcell swimmer with the swimming stroke Eq.~\eqref{comp} for different models. $a$, $b$) the exact solution determined by Eq.~\eqref{evo}. $c$, $d$) the minimal perturbation coordinate approximation. $e$, $f$) the embedded method using the exact path for $\theta(t)$. $g$, $h$) the embedded method using the guess path $\theta_g(t) =t \Delta \theta / 2 \pi$. } } 
\label{fig:compare}
\end{figure} 

}
\section{Visualisation of high dimensional swimmers}  \label{sec:vis}
  
 The previous section showed that the net displacement from any swimmer, in any background environment, can be represented exactly by a closed path integral in a higher dimensional space in which non-commuting variables are treated as prescribed paths. This representation allows the generalized Stokes' theorem to be applied, and so the net displacement can be related to the flux of a field through the path. The visualisation of this field can assist with the design of swimming stokes for specific tasks \citep{Keaveny2013,Quispe2019,Koens2018a} but is typically hard to do if there is more than three dimensions.  In this section we consider the behaviour of the net displacement throughout the entire embedded-configuration space of an arbitrary swimmer when the generalized Stokes theorem applies. We show that, if the embedded-space has more than two dimensions, every net displacement can be produced by an infinite set of swimming strokes. Different and equivalent swimming strokes can be visualised from this idea and the method is applied to a four-mode spherical squirmer { and the Purcell swimmer}.

\subsection{The divergence of the displacement field} \label{sec:dis3}

If the Stokes theorem applies, the net displacement of the swimmer in any direction can be written as
\begin{equation}
 \Delta x_{i} =\oint_{\partial V}M_{ij}  d l^{j} =    \iint_{V}  \partial M_{jk}^{i} d l^{j} \wedge d l^{k},
\end{equation}
where $V$ is any surface bounded by $\partial V$ and $\partial M_{jk}^{i} = -\frac{1}{2}\displaystyle\left( \frac{\partial M_{ij}}{\partial l^{k}} -\frac{\partial M_{ik}}{\partial l^{j}}  \right)$. The right-hand side of the above equation can be written as a vector product of the unique elements of $\partial \mathbi{M}^{i} \equiv \partial M_{jk}^{i}$ with the different infinitesimal surface elements $d l^{j} \wedge d l^{k}$. Hence this integral can be interpreted as the flux of  $\partial \mathbi{M}^{i}$ through any surface $V$ which is bounded by $\partial V$. This means the field $\partial \mathbi{M}^{i}$ must be divergence free. This is a consequence of the fact closed forms are exact in exterior calculus (see Appendix~\ref{sec:extior}). 
 This provides the displacement field with a strong parallel to incompressible fluid flow. In particular, there exists no one loop that can maximise the flux for systems with more than two modes of deformation ($N>2$). If a maximum displacement exists, there must be an infinite number of deformation loops that produce it.  These strokes may correspond to different environments in the embedded-configuration space.

 \begin{figure}
\centering
\includegraphics[width=0.6\textwidth]{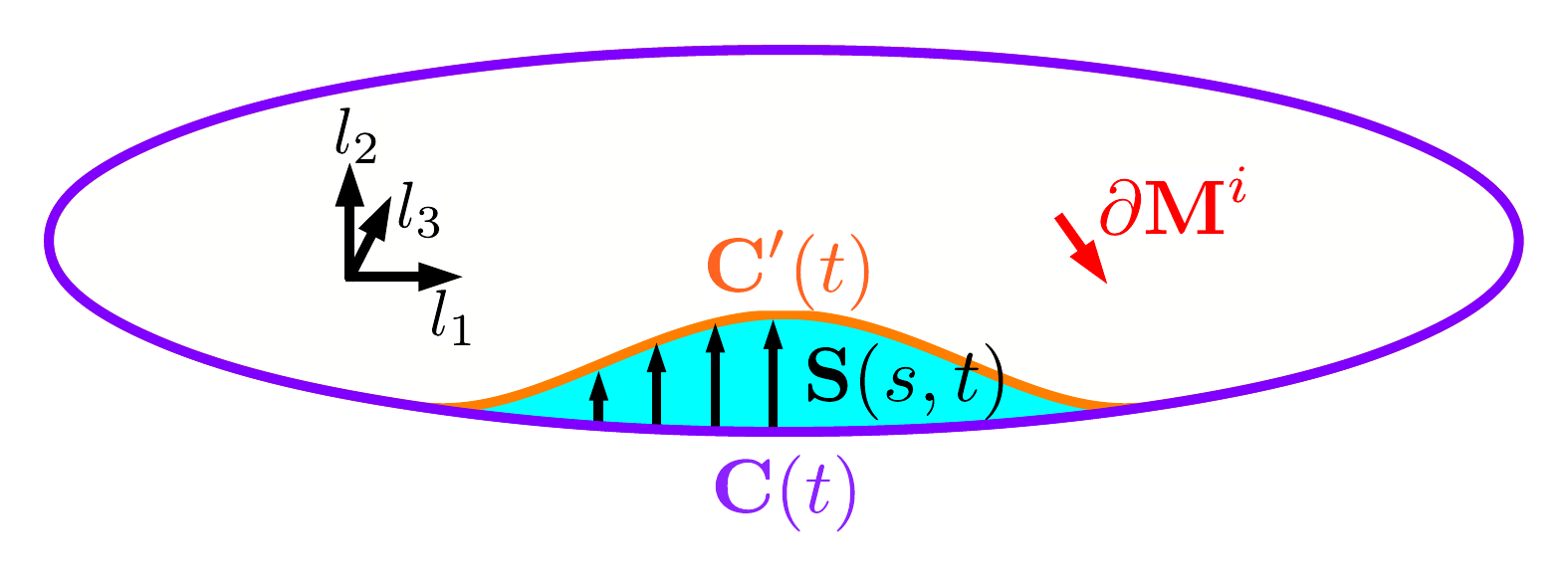}
\caption{Diagram depicting the deformation of a loop in a space with three modes of deformation. $\mathbf{C}(t)$ (dark purple) is the undeformed loop, $\mathbf{C}'(t)$ (orange) is the deformed loop, and the area in light blue is the surface $\mathbf{S}(s,t)$ created between the two loops. The arrows in the blue region point in the direction of $s$ and the red arrow is an example direction for $\partial \mathbi{M}^{i}$. } 
\label{fig:deflect}
\end{figure} 

\subsection{Non-equivalent strokes}  \label{sec:dis4}

The divergence-free nature of the field $\partial \mathbi{M}^{i}$ means that the displacement behaviour in one region can be `advected' into another. Hence, though the full space is high dimensional, not all of that space contains unique information. Two loops with the same displacement will be referred to as  equivalent strokes since they contain the same $\partial \mathbi{M}^{i}$ information, while two loops with different displacements will be said to be non-equivalent strokes. All the unique information in the field is therefore contained within a set of non-equivalent strokes that contains at least one stroke from every possible displacement. { To the best of our knowledge, the identification of these sets of non-equivalent strokes has not been considered before. We show that certain realisations of this} set can be constructed by considering how the flux through a stroke changes with an infinitesimal deformation { and using the results to form special surfaces in the configuration space. Similar flux catching techniques have been successfully used in several fields including electromagnetism and fluid flows.}  

 Consider a swimming stroke described in deformation space by $\mathbf{C}(t)$ where $t$ is the parametrisation around the loop. A deformation anywhere along this loop produces then a new loop given by $\mathbf{C}'(t)$ and the surface bridging these two loops is given by $\mathbf{S}(s,t)$  where $s$ is the parametrisation in the direction of the deflection (Fig.~\ref{fig:deflect}). The flux of $\partial \mathbi{M}^{i}$ through this surface is then given by
\begin{eqnarray}
\iint \partial M_{jk}^{i}  \frac{\partial S_{j}}{\partial s}  \wedge  \frac{\partial S_{k}}{\partial t} \,d s \,d t &=&  \iint \partial M_{jk}^{i}   \frac{\partial S_{j}}{\partial s}  \wedge  \frac{\partial S_{k}}{\partial t} \,d s \,d t \notag \\
&=&  \iint \partial M_{jk}^{i} \left(\frac{\partial S_{j}}{\partial s}  \frac{\partial S_{k}}{\partial t} - \frac{\partial S_{k}}{\partial s}  \frac{\partial S_{j}}{\partial t}\right) \,d s \,d t \notag \\
 &=& 2  \iint \partial M_{jk}^{i} \frac{\partial S_{j}}{\partial s}  \frac{\partial S_{k}}{\partial t} \,d s \,d t  \notag \\ 
 &\equiv & 2 \iint  \frac{\partial \mathbf{S}}{\partial s} \cdot \partial\mathbi{M}^{i} \cdot \frac{\partial \mathbf{S}}{\partial t}  \,d s \,d t, \label{sfull}
\end{eqnarray}
where we have used the asymmetric nature of the field, $\partial M_{jk}^{i} = -\partial M_{kj}^{i}$. In the limit $\,ds \to 0$, the above equation is a parametrised realisation of the Leibniz's rule \citep{Ramasamy2019}. The change in flux due to an infinitesimal displacements at any point along the loop is therefore related to
\begin{eqnarray}
 \Delta F= \frac{\partial \mathbf{S}}{\partial s} \cdot \partial\mathbi{M}^{i} \cdot \frac{\partial \mathbf{S}}{\partial t}.
\end{eqnarray}
where $\Delta F$ is the change in the flux per unit area. Since $\partial \mathbf{S}/\partial t= \partial \mathbf{C}/\partial t$ at the undisturbed loop, $\Delta F$ identifies how the displacement changes when a stroke is infinitesimally distorted in different directions. If the direction of $\partial \mathbf{S}/\partial s$ is perpendicular to  $ \partial\mathbi{M}^{i} \cdot (\partial \mathbf{S}/\partial t)$  the infinitesimal deformation generates no additional flux and so the strokes will be equivalent. However if $\partial \mathbf{S}/\partial s$ is parallel to $\partial\mathbi{M}^{i} \cdot (\partial \mathbf{S}/\partial t)$ the strokes can be different. 

In a $N$-dimensional configuration space, $\Delta F$ shows that { the direction, $\partial\mathbi{M}^{i} \cdot (\partial \mathbf{S}/\partial t)$, is responsible for the change in flux through a loop and the size of this change is linearly proportional to the amount of the distortion in this direction. Changes to the stroke which do not have a component in this direction (i.e. distortions which are a linear combination of the other $N-1$ directions) do not change the flux through the loop. {\color{blue} Hence any distortion containing a component in the $\partial\mathbi{M}^{i} \cdot (\partial \mathbf{S}/\partial t)$ direction will increase the net displacement of the stroke.} We note that the direction $\partial\mathbi{M}^{i} \cdot (\partial \mathbf{S}/\partial t)$ exists within the configuration space of the swimmer and so differs to the gradient of flux terms in \citet{Ramasamy2017,Ramasamy2019} which represent a direction in the parametrisation space of the loop itself. This gradient of flux representation can be derived from $\partial\mathbi{M}^{i} \cdot (\partial \mathbf{S}/\partial t)$ by multiplying $\partial\mathbi{M}^{i} \cdot (\partial \mathbf{S}/\partial t)$ by the distortion produced by changing each parametrisation mode of the loop and then integrating over the entire loop.}

{ The identification of this flux changing direction in the configuration space allows us to construct surfaces which contain the full set of non-equivalent strokes available to a swimmer. These surfaces, which we call surfaces of non-equivalent strokes, can be formed by distorting the loop in a direction that always contains a non-zero component in $\partial\mathbi{M}^{i} \cdot (\partial \mathbf{S}/\partial t)$. This ensures that as the loop distorts the flux through it changes. Recalling that $\partial \mathbf{S}/\partial s$ corresponds to the direction of distortion, the surfaces of non-equivalent strokes must satisfy a partial differential equation of the form }
\begin{equation}
\frac{\partial \mathbf{S}}{\partial s} = a(\mathbf{S}) \partial\mathbi{M}^{i}(\mathbf{S}) \cdot \frac{\partial \mathbf{S}}{\partial t} {+\mathbf{v}\left(\mathbf{S}, \frac{\partial \mathbf{S}}{\partial t}\right)}, \label{surf}
\end{equation}
where $a(\mathbf{S})$ is an arbitrary function of $\mathbf{S}$, { $\mathbf{v}\left(\mathbf{S}, \frac{\partial \mathbf{S}}{\partial t}\right)$ is an arbitrary vector orthogonal to $\partial\mathbi{M}^{i} \cdot (\partial \mathbf{S}/\partial t)$}, and we have included the dependence on $\mathbf{S}$ in $\partial\mathbi{M}^{i}(\mathbf{S})$ to be explicit. Different choices of $a(\mathbf{S})$ and {$\mathbf{v}\left(\mathbf{S}, \frac{\partial \mathbf{S}}{\partial t}\right)$} produce different surfaces of non-equivalent strokes, so it  can be freely chosen to simplify the governing equations. { Yet, even with this freedom, there are infinity many surfaces that do not satisfy this equation.} Throughout we will typically set $a(\mathbf{S}) =1$ and { $\mathbf{v}\left(\mathbf{S}, \frac{\partial \mathbf{S}}{\partial t}\right)=0$ }but note that if $\partial\mathbi{M}^{i}(\mathbf{S}) = \mathbf{0}$ { at specific values of $\mathbf{S}$, $a(\mathbf{S})$}  should be chosen such that $a(\mathbf{S}) \partial\mathbi{M}^{i}(\mathbf{S}) \neq \mathbf{0}$ everywhere. { If $\partial\mathbi{M}^{i}(\mathbf{S}) = \mathbi{0}$ everywhere, this is the trivial case and so the swimmer does not generate net displacement.} The parametrisation of the above equation is such that $s$ represents the different loops while $t$ is the position around the loop.  Importantly, for general $a(\mathbf{S})$, $s$ is not linked to a physical property of the selected loop and so it will depend on the initial loop chosen.

Since the displacement from a swimmer in any environment can be written as a closed path integral in the embedded-configuration space (Sec.~\ref{sec:commute}) and surfaces which satisfy Eq.~\eqref{surf} capture all non-equivalent strokes, the displacement from a swimmer, regardless of the environment, can always be visualised on a single surface. This is similar to the one-dimensional two-mode swimmer whose displacement can be visualised through a plane and so offers similar design possibilities. On this general surface the displacement from a specific environments exist over a continuous region of $s$, with maximal and minimal displacements for the system lying at the boundaries of these regions. If the region is closed, there must be at least one stroke which generates these maximal and minimal displacements, though there may be an infinite number of strokes. These regions are case- and surface-dependant and so their identification has been left for future work.

{\color{blue} Finally we note that, for a specific problem, it is possible to restrict the surfaces to only contain relevant strokes through the application of a projection operator. If the given system contains $N$ modes of deformation and $M$ non-commuting variables these constrained surfaces can be shown to satisfy
\begin{equation}
\frac{\partial \mathbf{S}}{\partial s} = \mathbf{P} \cdot \left[a(\mathbf{S}) \partial\mathbi{M}^{i}(\mathbf{S}) \cdot \frac{\partial \mathbf{S}}{\partial t} {+\mathbf{v}\left(\mathbf{S}, \frac{\partial \mathbf{S}}{\partial t}\right)}\right], \label{project}
\end{equation}
where the projection operator, $\mathbf{P}$, ensures that distortions are in the relevant directions and can be written as
\begin{equation}
\mathbf{P}  = \left(\begin{array}{c c}
\mathbf{I}_{N\times N} & \mathbf{0}_{N\times M} \\
\mathbf{M}(\mathbf{S}) & \mathbf{0}_{M\times M}
\end{array} \right).
\end{equation}
In the above $\mathbf{I}_{N\times N}$ is the identity matrix of size $N\times N$, and $\mathbf{0}_{i\times j}$ is a matrix of zeros of size $i\times j$.
  Though the above representation is appealing a priori, the surfaces from Eq.~\eqref{project} are no longer guaranteed to capture the full set of non-equivalent strokes available to the system and may in general miss the maximal displacement possible. Hence it is not generally possible to visualise all possible displacements from a swimmer on one of these surfaces, unlike the surfaces constructed from Eq.~\eqref{surf}.
}

\subsection{Equivalent strokes} \label{sec:dis5}

The non-divergent nature of $\partial\mathbi{M}^{i}$ implies that, if Eq.~\eqref{surf} can be solved, all other loops in the configuration space must be equivalent to at least one of the loops on $\mathbf{S}(s,t)$. This means that, similarly to the two-mode case, the displacement generated from any stroke in the space can be visualised through a single surface and so again provides an easy way to investigate the possible displacements generated. The design of strokes now corresponds to choosing different values of $s$ and searching for equivalent structures.

Equivalent strokes can be identified by considering different solutions to Eq.~\eqref{surf}. Since the parametrisation $s$ is not generally linked to a physical property of the loop, the parametrisation of each surface of non-equivalent loops found through Eq.~\eqref{surf} depends on the initial stroke used.  However, provided each surface spans all the net displacements possible, different surfaces must be composed of equivalent strokes  but with potentially different labels for $s$. This is due to freedom in the choice of parametrisation for the surface. For example, consider two surfaces $\mathbf{S}_a(s,t)$ and $\mathbf{S}_b(s',t')$. If the $\mathbf{S}_a(s=1,t)$ loop generates the same displacement as  to the $\mathbf{S}_b(s'=5,t')$ loop, the two loops are equivalent but have been labelled differently. Since the parametrisation of surfaces are not unique, the identification equivalent loops is therefore the same as rescaling the $s$ parametrisation on the different surfaces in relation to the net displacement through each loop. This can be achieved by considering the total change in the flux with $s$ which is given by
\begin{equation}
\frac{d \Delta x_{i}}{d s} = \oint \frac{\partial \mathbf{S}}{\partial s}\cdot \partial\mathbi{M}^{i} \cdot \frac{\partial \mathbf{S}}{\partial t} \,d t, \label{cdx}
\end{equation}
where the integral has been taken over the full loop. This equation determines the displacement of every loop on the surface simultaneously and so removes the need to compute several loops individually to construct a similar space. If two solutions to Eq.~\eqref{surf} generate the same Eq.~\eqref{cdx}, then the equivalent loops must already share the same $s$ to within a constant shift. If however they generate different Eq.~\eqref{cdx}, the $s$ on one of the solutions can be rescaled such that they are the same. This could be achieved by writing the loops in terms of $\Delta x_{i}$ instead of $s$.  Importantly this re-parametrisation process is closely linked to the transformation invariants of Eqs.~\eqref{surf} and \eqref{cdx}. If both these equations are invariant to translation, then any translation of the original solution would produce a set of equivalent loops.

\subsection{Example: Four-mode Squirmer}

 \begin{figure}
\centering
\includegraphics[width=0.9\textwidth]{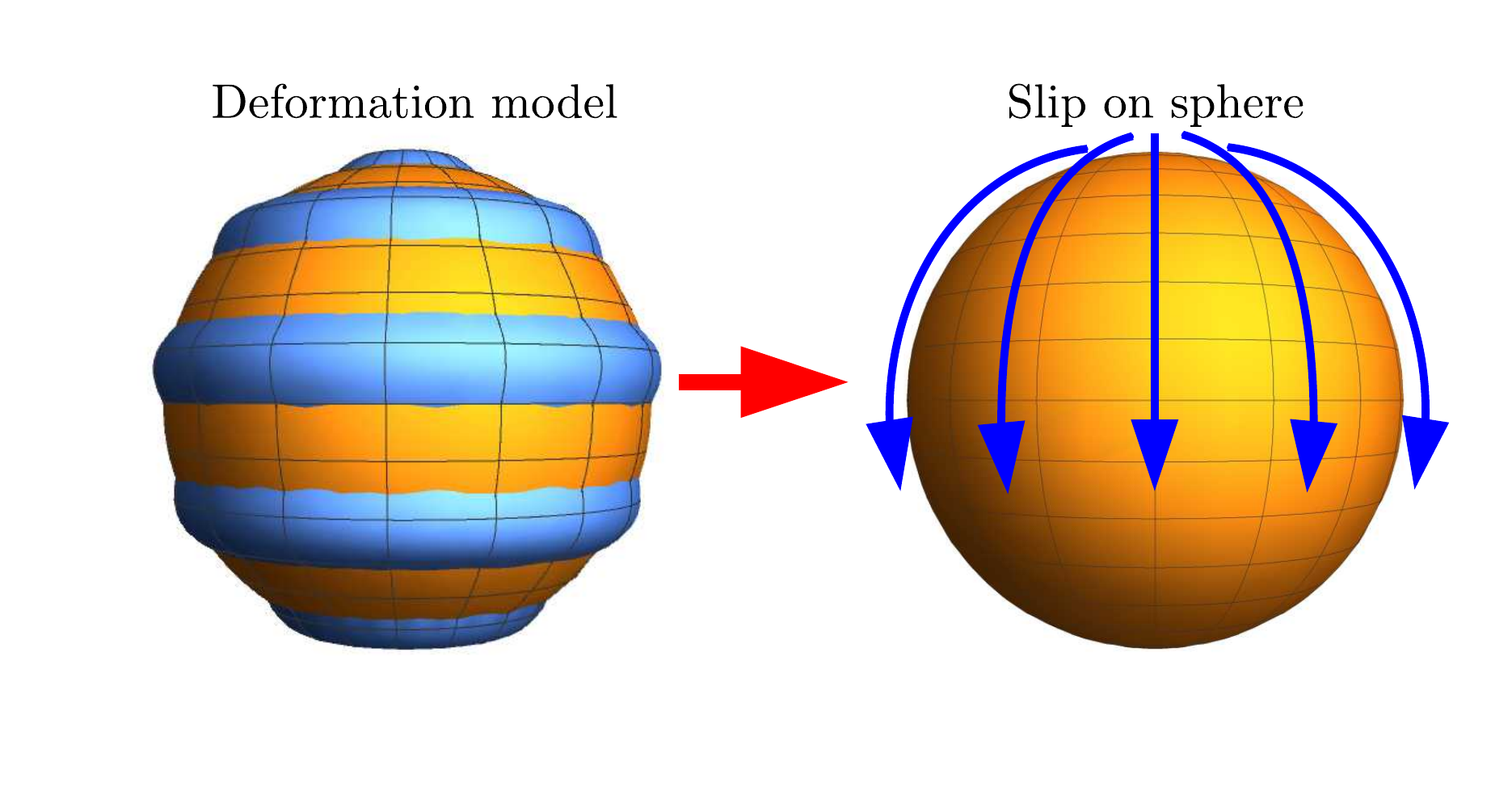}
\caption{Diagram depicting a deforming sphere and a perfect sphere with slip. In the small deformation limit, the flow around a deforming sphere can be treated as a sphere with effective slip boundary conditions.} 
\label{fig:squirm1}
\end{figure}

The  prototypical swimmer with multiple discrete modes of deformation is the spherical squirmer. Squirmers are spherical bodies that periodically deform their surface in order to generate motion.  These surface deformations are typically assumed to be small and so can be expanded as an effective slip velocity over its surface (Fig.~\ref{fig:squirm1}). This swimmer was first proposed by \citet{Lighthill1952}, the calculation was corrected by \citet{Blake1971a}, and has since had many extensions which consider both the swimmers speed and efficiency \citep{Shapere1989, Shapere1989a,Ishimoto2013,Pedley2016,Felderhof2016a,Eastham2019}. Notably \citet{Shapere1989} used the geometric swimming techniques to derive an equation for the mean velocity in the limit of small, but arbitrary, surface deformations. However this form is too complicated for demonstration and so the original axisymmetric model of Blake will be employed. The motion of this squirmer does not induce any non-commuting variables but is complex enough for the equivalent/non-equivalent stroke techiques described above to be of use.

\subsubsection{Squirmer formulation}

The oscillating surface of an axisymmetric squirmer \citep{Blake1971a} can be parametrised by the polar surface angle of the base sphere, $\theta_{0}$, as
\begin{eqnarray}
R &=& 1 + \epsilon  \sum_{n=2}^{N} \alpha_{n}(t) P_{n}(\cos\theta_{0}) , \\
\theta &=& \theta_{0} +  \epsilon  \sum_{n=1}^{N} \beta_{n}(t) V_{n}(\cos\theta_{0}) ,
\end{eqnarray}
where $(R,\theta)$ is the radial and polar position of the deformed sphere sphere, $P_{n}(x)$ is the Legendre polynomial of order $n$, $V_{n}(\cos\theta_{0}) =2 \sin \theta_{0} P_{n}'(\cos\theta_{0})/(n(n+1))$, $\epsilon$ is small, $2N-1$ is the number of modes and $\alpha_{n}(t) $ and $\beta_{n}(t)$ are periodic functions of time (Fig.~\ref{fig:squirm1}). The swimming velocity of this squirmer to order $\epsilon^{3}$ is \citep{Blake1971a}
\begin{eqnarray}
U &=&  \frac{2\epsilon}{3}  \dot{\beta}_{1} - \frac{8\epsilon^{2}}{15} \alpha_{2} \dot{\beta}_{1} - \frac{2\epsilon^{2}}{5} \dot{\alpha}_{2} \beta_{1} + \epsilon^{2} \sum_{n=1}^{N-1}\frac{4(n+2)\beta_{n}\dot{\beta}_{n+1} - 4 n \dot{\beta}_{n}\beta_{n+1}}{(n+1)(2n+1)(2n+3)} \notag \\
&& + \epsilon^{2}\sum_{n=2}^{N-1}\frac{(2n+4)\alpha_{n}\dot{\beta}_{n+1} -2 n   \dot{\alpha}_{n}\beta_{n+1} - (6n+4) \alpha_{n+1} \dot{\beta}_{n} - (2n+4) \dot{\alpha}_{n+1} \beta_{n}}{(2n+1)(2n+3)} \notag \\
&& -\epsilon^{2} \sum_{n=2}^{N-1}\frac{(n+1)^2\alpha_{n}\dot{\alpha}_{n+1} - (n^{2}-4n-2) \dot{\alpha}_{n}\alpha_{n+1}}{(2n+1)(2n+3)} ,
\end{eqnarray}
where we use $\dot{(\cdot)}$ to mean  derivatives with respect to time. The displacement from this velocity can be treated generally with the geometric swimming techniques. Here we will set $\alpha_{n}=0$ and $N=4$ for demonstration purposes, in which case the the velocity becomes
\begin{eqnarray}
U &=& \left( \frac{2\epsilon}{3} -  \frac{2\epsilon^{2}}{15}\beta_{2}\right) \dot{\beta}_{1} + \frac{2 \epsilon^{2}}{105} \left( 21 \beta_{1} - 4 \beta_{3}\right)\dot{\beta}_{2} + \frac{ \epsilon^{2}}{105} \left( 16\beta_{2} - 5 \beta_{4}\right)\dot{\beta}_{3} + \frac{5\epsilon^{2}}{63}\beta_{3} \dot{\beta}_{4} \notag \\
&=& \left\{\frac{2\epsilon}{3} -  \frac{2\epsilon^{2}}{15}\beta_{2}, \frac{2 \epsilon^{2}}{105} \left( 21 \beta_{1} - 4 \beta_{3}\right),\frac{ \epsilon^{2}}{105} \left( 16\beta_{2} - 5 \beta_{4}\right), \frac{5\epsilon^{2}}{63}\beta_{3} \right\} \cdot \frac{d}{dt}\left\{\beta_{1},\beta_{2},\beta_{3},\beta_{4} \right\} . \notag \\
\end{eqnarray}

The net displacement, without using the generalized Stokes theorem, may thus be written as 
\begin{equation}
\Delta x =  \oint \mathbf{M}^{Sq}\cdot d \boldsymbol{l}^{Sq},
\end{equation}
where the integral is taken over a specified path and
\begin{eqnarray}
\mathbf{M}^{Sq} &=& \left\{\frac{2\epsilon}{3} -  \frac{2\epsilon^{2}}{15}\beta_{2}, \frac{2 \epsilon^{2}}{105} \left( 21 \beta_{1} - 4 \beta_{3}\right),\frac{ \epsilon^{2}}{105} \left( 16\beta_{2} - 5 \beta_{4}\right), \frac{5\epsilon^{2}}{63}\beta_{3} \right\}, \\
d\boldsymbol{l}^{Sq} &=& \left\{d\beta_{1},d\beta_{2},d\beta_{3},d\beta_{4} \right\}.
\end{eqnarray}
The behaviour of the net displacements can be found from the above equation by prescribing different loops and exploring its dependence on them. 

However it is easy to misinterpret these results. Consider for example  the three specific  strokes $\mathbf{C}^{1}(t) = \{ \cos t, \sin t ,0,0\}$, $\mathbf{C}^{2}(t) = \{ \cos t,0 , \sin t, 0\}$, and $\mathbf{C}^{3}(t) = \{ 0, 0,  \cos t , 21 \sin t /5 \}$. The first and second of these loops represent circles in the $\beta_{1}-\beta_{2}$ and $\beta_{1}-\beta_{3}$ planes respectively, while the last is an ellipsoid in the $\beta_{3}-\beta_{4}$ plane with semi-axes lengths $1$ and $21/5$.  The net displacements from each of these loops can be  integrated directly to find
\begin{eqnarray}
\Delta x[\mathbf{C}^{1}(t)] &=& \frac{8 \pi \epsilon^{2}}{15}, \\
\Delta x[\mathbf{C}^{2}(t)] &=& 0, \\
\Delta x[\mathbf{C}^{3}(t)] &=& \frac{8 \pi \epsilon^{2}}{15}.
\end{eqnarray}
 These results indicate that the displacement of a squirmer depends strongly on the shape and the position of the loop in the configuration space. For example, though $\mathbf{C}^{1}(t)$ and $\mathbf{C}^{2}(t)$ are both circles, they do not share the same displacement because the correspond the very different strokes. However this is not the full story as $\mathbf{C}^{1}(t)$ and $\mathbf{C}^{3}(t)$ also correspond to very different strokes but generate the same displacement. Hence if we wanted to find a relationship between these loops, we would (normally) need to study many more strokes to capture the full picture.

\subsubsection{Properties of the displacement field}

Using our new approach, the displacement field of the squirmer allows us to rationalise  these results. The generalized Stokes theorem allows us to write the displacement as
\begin{equation}
\Delta x =  \oint \mathbf{M}^{Sq}\cdot d \boldsymbol{l}^{Sq} = \iint \partial\mathbf{M}^{Sq} :  \left(d \boldsymbol{l}^{Sq} \wedge d \boldsymbol{l}^{Sq}\right),
\end{equation}
where $:$ denotes the double contraction. Here the path integral is to be taken over a loop, the area integral is taken over any surface bounded by said loop and we have
\begin{eqnarray}
\partial\mathbf{M}^{Sq} &=& 4 \epsilon^{2}\left(\begin{array}{c c c c}
0 & -\frac{1}{15} & 0&0  \\
\frac{1}{15}& 0& -\frac{1}{35}& 0 \\
0& \frac{1}{35}&0 &-\frac{1}{65}  \\
0& 0& \frac{1}{65}&0
\end{array} \right),\\
d \boldsymbol{l}^{Sq} \wedge d \boldsymbol{l}^{Sq} &=& \left(\begin{array}{c c c c}
0 & d\beta_{1}\wedge d \beta_{2} & d\beta_{1}\wedge d \beta_{3}  & d\beta_{1}\wedge d \beta_{4}  \\
-d\beta_{1}\wedge d \beta_{2}  & 0 &d\beta_{2}\wedge d \beta_{3} & d\beta_{2}\wedge d \beta_{4} \\
-d\beta_{1}\wedge d \beta_{3} & d\beta_{2}\wedge d \beta_{3} & 0 &d\beta_{3}\wedge d \beta_{4} \\
- d\beta_{1}\wedge d \beta_{4} &  -d\beta_{2}\wedge d \beta_{4} & - d\beta_{3}\wedge d \beta_{4}& 0
\end{array} \right).
\end{eqnarray}
The off-diagonal structure of $\partial\mathbf{M}^{Sq}$ is standard for systems in which only the nearest neighbour modes are coupled while the constant coefficients are typical for systems expanded to leading order in shape non-linearities. The surfaces to integrate over for each of the example strokes can  be easily produced by simply multiplying each curve by $s \in [0,1]$, and the results found are identical to the displacements above.
This matrix representation for the displacement can be simplified further through the coordinate transformation
\begin{eqnarray}
 x_{1} &=& \cos \theta_{1}   \beta_{1} + \sin \theta_{1} \beta_{3}, \\
 x_{2} &=& \cos \theta_{2}   \beta_{2} + \sin \theta_{2}  \beta_{4}, \\
 x_{3} &=& -\sin \theta_{1}   \beta_{1} + \cos \theta_{1}  \beta_{3}, \\
 x_{4} &=& -\sin \theta_{2}   \beta_{2} + \cos \theta_{2}  \beta_{4},
\end{eqnarray}
where $\theta_{1} = \arctan\left( \frac{\sqrt{255109}-335}{378}\right)$ and $\theta_{2} = \arctan\left( \frac{\sqrt{255109}-497}{90}\right)- \pi$. In these coordinates, $\partial\mathbf{M}^{Sq}$ becomes
\begin{equation}
\partial\mathbf{M}^{Sq}_{R} = \frac{ 2\epsilon^{2}}{315} \left(\begin{array}{c c c c}
0 & \sqrt{337} + \sqrt{757} & 0&0  \\
-\sqrt{337} -\sqrt{757} & 0& 0& 0 \\
0& 0&0 &-\sqrt{337} + \sqrt{757}  \\
0& 0& \sqrt{337} - \sqrt{757}&0
\end{array} \right).\\
\end{equation} 
and the net displacement can be written as
\begin{eqnarray}
\Delta x &=&  \iint \partial\mathbf{M}^{Sq}_{R} :  \left(d \boldsymbol{x} \wedge d \boldsymbol{x}\right) \notag \\
&=& \frac{ 4\epsilon^{2}}{315}\iint \left(\sqrt{337} + \sqrt{757} \right) dx_{1} \wedge dx_{2} + \left( \sqrt{757}-\sqrt{337}  \right) dx_{3} \wedge dx_{4}. \label{spdx}
\end{eqnarray}
A similar coordinate transformation is always possible for constant $\partial\mathbf{M}^{Sq}$ matrices but is not always possible for general $\partial\mathbf{M}^{Sq}$ (Appendix~\ref{sec:skew}). This representation of the displacement indicates that only the areas enclosed by a stroke in the $x_1-x_2$ and $x_3-x_4$ planes actually contribute to the net displacement. These contributions are scaled by a specific factor related to the matrix eigenvalues (Appendix~\ref{sec:skew}). 

For a $N$-mode system with a constant $\partial\mathbf{M}^{Sq}$, this representation for the net displacement extends  to
\begin{equation}
\Delta x = 2 \lambda_{1} \iint d x^{1} \wedge d x^{2}  + 2 \lambda_{2} \iint d x^{3} \wedge d x^{4}  + \cdots 2 \lambda_{n} \iint d x^{2n-1} \wedge d x^{2n} + \cdots \label{spdxN}
\end{equation}
where $\lambda_{i}$ are positive constants. If $N$ is even there are $N/2$ non-zero values of $\lambda_{i}$, while if $N$ is odd there are $(N-1)/2$ non-zero values. The above is a direct result of the existence of a the aforementioned coordinate transformation. We note that locally such a transformation exists for any $\partial\mathbf{M}^{Sq}$ (see Appendix~\ref{sec:skew}) and so Eq.~\eqref{spdxN} is always locally true. 

 In these coordinates, the example loops become
\begin{eqnarray}
\mathbf{C}^{1}(t) &=& \{\cos \theta_{1} \cos t, \cos \theta_{2} \sin t, - \sin \theta_{1} \cos t,- \sin \theta_{2} \sin t\}, \\
\mathbf{C}^{2}(t) &=& \{\cos \theta_{1} \cos t + \sin \theta_{1} \sin t, 0,\cos \theta_{1} \sin t - \sin \theta_{1} \cos t,0\}, \\
\mathbf{C}^{3}(t) &=& \{ \sin \theta_{1}   \cos t, \frac{21}{5} \sin \theta_{2} \sin t,  \cos \theta_{1} \cos t,\frac{21}{5}  \cos \theta_{2} \sin t \}.
\end{eqnarray}
The second stroke, $\mathbf{C}^{2}(t)$, only has components in the $x_{1}$ and $x_{3}$ direction. Hence it encloses no area in the $x_1-x_2$ or $x_3-x_4$ planes and so its displacement must be $0$, as we had obtained. This coordinate representation therefore identifies why $\mathbf{C}^{2}(t)$ generates no displacement by inspection. The same idea can be applied to many other loops.

\subsubsection{Non-equivalent strokes}

The behaviour of the field around any loop can be interpreted through a surface of non-equivalent strokes created from the loop. These surfaces reveal the direction in which the loops change value and identify the remaining space as equivalent. We will look for surfaces that satisfy Eq.~\eqref{surf} when $a(\mathbf{S})=1$ and $\mathbf{v}=0$. In the rotated coordinates this equation becomes
\begin{eqnarray}
\frac{\partial S_{1}}{\partial s} &=& \frac{ 2\epsilon^{2}}{315} \left(\sqrt{337} + \sqrt{757} \right)\frac{\partial S_{2}}{\partial t}, \\
\frac{\partial S_{2}}{\partial s} &=& -\frac{ 2\epsilon^{2}}{315} \left(\sqrt{337} + \sqrt{757} \right)\frac{\partial S_{1}}{\partial t}, \\
\frac{\partial S_{3}}{\partial s} &=& \frac{ 2\epsilon^{2}}{315} \left(\sqrt{757}-\sqrt{337} \right) \frac{\partial S_{4}}{\partial t}, \\
\frac{\partial S_{4}}{\partial s} &=& -\frac{ 2\epsilon^{2}}{315} \left(\sqrt{757}-\sqrt{337}  \right) \frac{\partial S_{3}}{\partial t},
\end{eqnarray}
where $\mathbf{S}(s,t) = \{ S_{1}(s,t), S_{2}(s,t), S_{3}(s,t), S_{4}(s,t)\}$, in the rotated frame. These equations rescale to
\begin{eqnarray}
\frac{\partial S_{1}}{\partial s_{1}} &=& \frac{\partial S_{2}}{\partial t}, \\
\frac{\partial S_{2}}{\partial s_{1}} &=& -\frac{\partial S_{1}}{\partial t}, \\
\frac{\partial S_{3}}{\partial s_{2}} &=&  \frac{\partial S_{4}}{\partial t}, \\
\frac{\partial S_{4}}{\partial s_{2}} &=& - \frac{\partial S_{3}}{\partial t},
\end{eqnarray}
where $s_{1} = \frac{ 2\epsilon^{2}}{315} \left(\sqrt{337} + \sqrt{757} \right) s$ and $s_{2} = \frac{ 2\epsilon^{2}}{315} \left(\sqrt{757}-\sqrt{337}  \right) s$. The equations above are two sets of the Cauchy–Riemann equations. Hence each $S_{i}$ must satisfy a two-dimensional Laplace equation and the general solution can be represented through two holomorphic functions as
\begin{equation}
\mathbf{S}(s,t) = \{\Re[f_{1}(s_{1}+ i t)],\Im[f_{1}(s_{1}+ i t)],\Re[f_{2}(s_{2}+ i t)],\Im[f_{2}(s_{2}+ i t)] \}, \label{gensol}
\end{equation}
where $\Re[\cdot]$ denotes the real part, $\Im[\cdot]$ denotes the imaginary part and $f_{1}(s+i t)$ and $f_{2}(s+i t)$ are arbitrary analytic functions in which $t$ describes a closed loop. This general solution set extends trivially to any $\partial\mathbf{M}^{Sq}$ which is a constant $N\times N$ matrix, with the solution being represented through $N/2$ or $(N-1)/2$ analytic functions if $N$ is even or odd respectively. 

The ability to write the general solution in terms of a collection of analytic functions is a reflection of the distinct planes which generate motion. In the case of the squirmer, $f_{1}(s_{1}+ i t)$ provides a parametrisation of the $x_1-x_2$ plane while $f_{2}(s_{2}+ i t)$ is the parametrisation of the $x_3-x_4$ plane. Conformal maps are just re-parametrisations of these planes. However since re-parametrisations of each plane can be done independently, the actual surfaces can look very different.

 \begin{figure}
\centering
\includegraphics[width=0.7\textwidth]{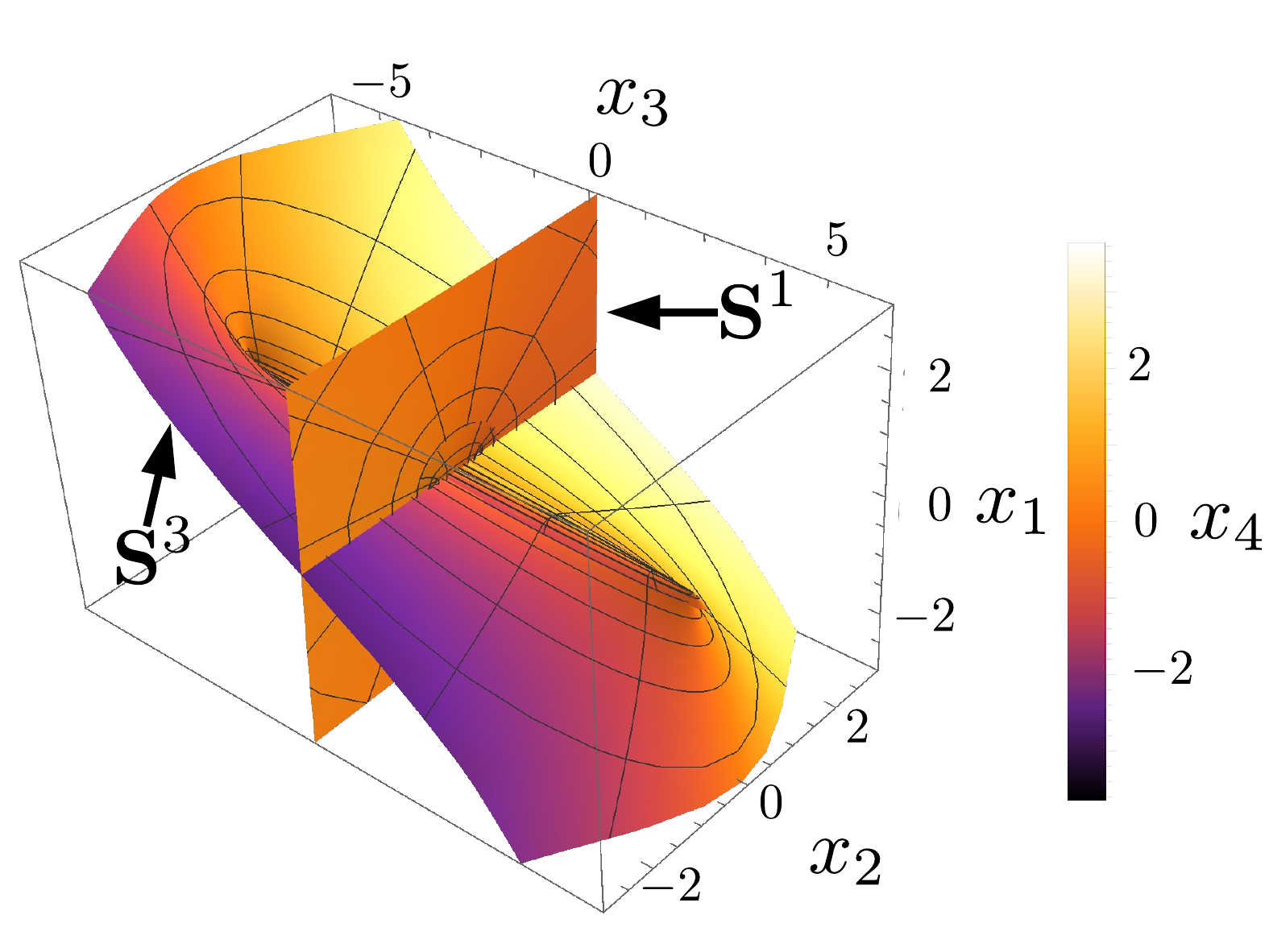}
\caption{Two surfaces of strokes that generate different displacements for four-mode squirmer.  $\mathbf{S}^{1}(s,t)$ is shown in orange and $\mathbf{S}^{3}(s,t)$ in light blue.  The color of the surface corresponds the its position in the forth dimension, $x_4$ } 
\label{fig:squirm2}
\end{figure} 

The application of boundary conditions to the general solution allows us to identify specific surfaces of loops with different displacements. For example the surfaces generated from $\mathbf{C}^{1}(t)$ and $\mathbf{C}^{3}(t)$ are
\begin{eqnarray}
\mathbf{S}^{1}(s,t) &=& \left\{\left(\cos \theta_{1} \cosh s_1 - \cos \theta_2 \sinh s_1 \right) \cos t, \left(\cos \theta_{2} \cosh s_1 - \cos \theta_1 \sinh s_1 \right) \sin t, \right. \notag \\ 
&& \left. \left(\sin \theta_1 \sinh s_2  - \sin \theta_{2} \cosh s_2  \right) \cos t,\left(\sin \theta_2 \sinh s_2  - \sin \theta_{1} \cosh s_2  \right) \sin t \right\}, \\
\mathbf{S}^{3}(s,t)
&=& \left\{\left( \sin \theta_{1}  \cosh s_1 - b \sin \theta_{2} \sinh s_1 \right) \cos t, \left( b \sin \theta_{2} \cosh s_1 - \sin \theta_{1} \sinh s_1 \right) \sin t, \right. \notag \\ 
&& \left. \left(b\cos \theta_{2} \cosh s_2   -  \cos \theta_{1} \sinh s_2 \right) \cos t,\left(   \cos \theta_{1} \cosh s_2  -b \cos \theta_{2} \sinh s_2  \right) \sin t \right\}, \notag \\
\end{eqnarray}
where $b= 21/5$, $\mathbf{S}^{1}(s,t)$ is the surface which becomes $\mathbf{C}^{1}(t)$ at $s=0$  and $\mathbf{S}^{3}(s,t)$ is the surface of loops which becomes $\mathbf{C}^{3}(t)$ at $s=0$. These surfaces are plotted in Fig.~\ref{fig:squirm2} and are visually very different. This is because the direction in which the deflection increases the flux, $\partial\mathbf{M}^{i} \cdot (\partial \mathbf{S}/\partial t)$, depends on the shape of the initial loop. These surfaces inherently identify a set of strokes of different displacement. All other loops are equivalent to one on the surface. The design of specific displacements is therefore reduced to exploring the displacement of specific strokes over the surface, rather than over a high-dimensional space. Once a desired displacement is identified, equivalent loops can then be considered.

\subsubsection{Equivalent strokes}

A solution to Eq.~\eqref{surf} contains, by construction, equivalent loops to another solution of Eq.~\eqref{surf} but with a different label for $s$. The identification of equivalent loops  requires therefore unifying these labels across the solutions.  Since two loops are equivalent if they produce the same displacement, this unification is possible through the change in net displacement with $s$ across a surface, Eq.~\eqref{cdx}. For the four-mode squirmer, Eq.~\eqref{cdx} is 
\begin{eqnarray}
\frac{d \Delta x}{d s} &=& \oint \left( \partial\mathbf{M}^{Sq}_{R} \cdot \frac{\partial \mathbf{S}}{\partial t}\right)^{2} \,d t  \notag \\
&=&  \left(\frac{ 2\epsilon^{2} \left(\sqrt{337} + \sqrt{757} \right)}{315}\right)^{2} \oint \left[ \left(\frac{\partial S_{1}}{\partial t} \right)^{2}+  \left(\frac{\partial S_{2}}{\partial t} \right)^{2} \right] \,d t\notag \\
&& +\left(\frac{ 2\epsilon^{2} \left( \sqrt{757}-\sqrt{337} \right)}{315}\right)^{2}\oint \left[ \left(\frac{\partial S_{3}}{\partial t} \right)^{2}+ \left(\frac{\partial S_{4}}{\partial t} \right)^{2} \right] \,d t \notag \\
&=&  \left(\frac{ 2\epsilon^{2} \left(\sqrt{337} + \sqrt{757} \right)}{315}\right)^{2} \oint \left|\frac{\partial f_{1}}{\partial t}\right|^{2} \,d t +\left(\frac{ 2\epsilon^{2} \left( \sqrt{757}-\sqrt{337} \right)}{315}\right)^{2}\oint  \left|\frac{\partial f_{2}}{\partial t}\right|^{2} \,d t \notag \\ \label{cdisp}
\end{eqnarray}
where $|\cdot|^{2}$ is the complex modulus. This equation can be used in many ways to identify the equivalent loops on different solutions  to Eq.~\eqref{surf}. The solution of Eq.~\eqref{cdisp} determines the net displacement of every loop on a surface as a function of the loop label parameter $s$. This allows the loops in said surface to be re-parametrised by $\Delta x$. If this process is repeated on each surface all equivalent loops will be labelled by the same $\Delta x$.

Alternatively, rather than finding the net displacement from every loop on each surface, Eq.~\eqref{cdisp} can also be used to re-parametrise one surface in terms of the $s$ variable of another surface. Consider the surfaces $\mathbf{S}^{a}(s,t)$ and $\mathbf{S}^{b}(s',t)$. The use of the chain rule on Eq.~\eqref{cdisp} shows that $s$ and $s'$ on these surfaces can be related by
\begin{eqnarray}
\oint \left( \partial\mathbf{M}^{Sq}_{R} \cdot \frac{\partial \mathbf{S}^a(s,t)}{\partial t}\right)^{2} \,d t  &=& \frac{d \Delta x}{d s}  \notag \\
&=& \frac{d \Delta x}{d s'} \frac{d s'}{d s}  \notag \\
 &=& \left[\oint \left( \partial\mathbf{M}^{Sq}_{R} \cdot \frac{\partial \mathbf{S}^b(s',t)}{\partial t}\right)^{2} \,d t  \right] \frac{d s'}{d s}. \label{equiv}
\end{eqnarray}
If we know a loop on each surface that generates the same displacement, this equation creates a relationship between $s'$ and $s$ such that the loop $\mathbf{S}^{b}(s'(s),t)$ is equivalent to the loop $\mathbf{S}^{a}(s,t)$. This relationship is illustrated with $\mathbf{S}^{1}(s,t)$ and $\mathbf{S}^{3}(s',t)$ in Fig.~\ref{fig:squirm3}a. This solution was found numerically using the condition that the flux is the same at $s'=s=0$. From this figure the equivalent loops in the two configurations and be easily identified. For example, we see that the loop $\mathbf{S}^{1}(s=-10,t)$ has the same displacement as $\mathbf{S}^{3}(s'\approx -7,t)$. This process allows the equivalence to be identified without needing to calculate $\Delta x$ for each surface, thereby reducing the computations needed. Furthermore if $\Delta x$ was known on one surface, the values of $\Delta x$ on the second can be easily deduced.

 \begin{figure}
\centering
\includegraphics[width=\textwidth]{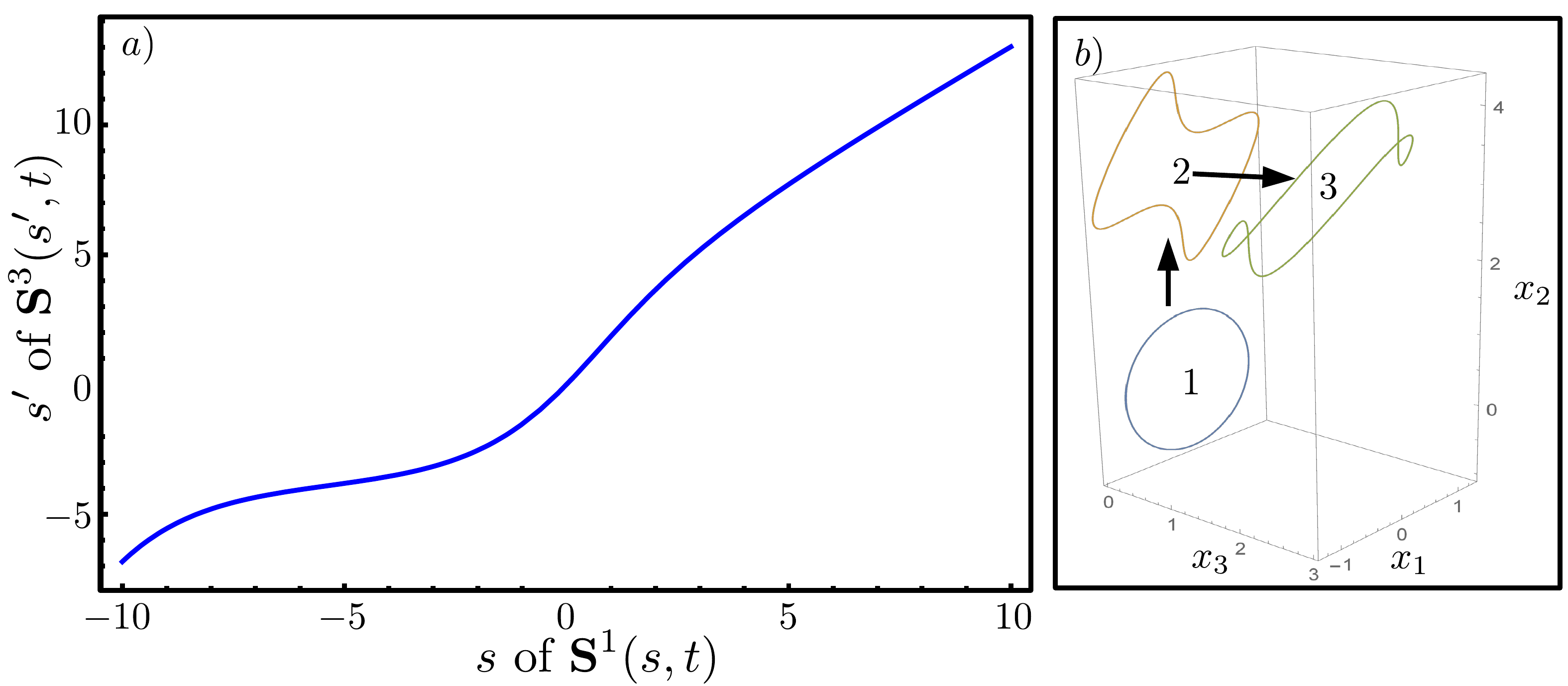}
\caption{a) The equivalence relationship between $s'$ and $s$ for $\mathbf{S}^{3}(s',t)$ and $\mathbf{S}^{1}(s,t)$. The relationship is structured such that the displacement from $\mathbf{S}^{3}(s'(s),t)$ is the same as $\mathbf{S}^{1}(s,t)$. b) Diagram of equivalent loop construction. Starting with loop 1 (blue) in the $x_{1}-x_2$ plane the loop is distorted into loop 2 (orange) shape that preserves the area but also sits in the $x_{1}-x_2$ plane. The loop is then transformed into loop 3 (green) by the equivalence relationship for projecting loops.} 
\label{fig:squirm3}
\end{figure} 

Equivalent loops can also be identified through the symmetries of Eq.~\eqref{surf}. Alterations of a surface of non-equivalent loops that leave Eqs.~\eqref{cdisp} and \eqref{surf} unchanged retain the same parametrisation of $s$ (to within a constant). This means that the equivalent loops are easily identified. Translations, rotations and reflections in the $x_{1}-x_{2}$ and $x_{3}-x_{4}$ planes of any solution are example transformations which retain the same parametrisation in the squirmer problem. A slightly more complicated transformation that also satisfies these conditions is
\begin{multline}
\{S_1(s_1,t) ,S_2(s_1,t),S_3(s_2,t),S_4(s_2,t) \} \mapsto \\ \{0,0,S_3(s_2,t)+\frac{1}{\sigma} S_1(\sigma s_1,t),S_4(s_2,t) +\frac{1}{\sigma} S_2(\sigma s_1,t) \}
\end{multline}
\begin{multline}
\{S_1(s_1,t) ,S_2(s_1,t),S_3(s_2,t),S_4(s_2,t) \} \mapsto  \\ \{S_1(s_1,t) +\sigma S_3(s_2/\sigma,t) ,S_2(s_1,t) + \sigma S_4(s_2/\sigma,t),0,0 \},
\end{multline}
where $\sigma = \left(\sqrt{757}-\sqrt{337}  \right) /\left(\sqrt{337} + \sqrt{757} \right)$. This transformation maps the contributions between the $x_{1}-x_{2}$ and the $x_{3}-x_{4}$ planes. This enables any surface to be visualised on the $x_{1}-x_{2}$ or $x_{3}-x_{4}$ plane alone. Since any stroke on a surface of non-equivalent strokes can always be parametrised such that $s=0$ this transformation also applies to any loop in the space.
For example this relationship states that all the loops in the $\mathbf{S}^{1}(s,t)$ surface are equivalent to 
\begin{eqnarray}
\mathbf{S}^{1}(s,t) &\mapsto&  \left\{\left(\cos \theta_{1} \cosh s - \cos \theta_2 \sinh s \right) \cos t + \sigma \left(\sin \theta_1 \sinh s  - \sin \theta_{2} \cosh s  \right) \cos t, \notag \right. \\ 
&& \quad \left(\cos \theta_{2} \cosh s - \cos \theta_1 \sinh s \right) \sin t + \sigma \left(\sin \theta_2 \sinh s  - \sin \theta_{1} \cosh s  \right) \sin t, \notag \\
&& \qquad 0,0 \}, \label{transforms}
\end{eqnarray}
where equivalent loops share the same parametrisation of $s$. The above transformation means that the net displacement of any of stroke, Eq.~\eqref{spdx},  can be determined from the area within the loop when mapped onto a suitable plane. This is because Eq.~\eqref{spdx} relates the displacement of a loop to the area enclosed within the $x_{1}-x_{2}$ or $x_{3}-x_{4}$ planes.  From this we see that the strokes $\mathbf{C}^{1}(t)$ and $\mathbf{C}^{3}(t)$ enclose the same area when mapped to the $x_{1}-x_{2}$ or $x_{3}-x_{4}$ plane and so generate the same displacement.  This ability to visualise any loop on a single plane and relate its displacement to the area enclosed extends to any system with constant $\partial\mathbf{M}^{Sq}$ but with different scaling factors. Hence, identically to the two-mode swimmers, all the loops can be visualised on a plane and the displacement determined from the integral within this loop.

Finally in the case where $\partial\mathbf{M}^{Sq}$ is constant, the inverse of these plane transformation maps also hold. The full space of equivalent loops can therefore be constructed from the set of loops on the $x_{1}-x_{2}$ plane that preserve the area within the stroke combined with all possible maps out of the surface that preserve the flux (Fig.~\ref{fig:squirm3}b). Equivalent loops could therefore be designed by transforming the loop on the plane through a method that conserves area and then projecting the loop out into the space.

{
 \subsection{Example: Surfaces for translation of Purcell swimmer}

Similarly to the squirmer, surfaces of non-equivalent strokes can also be constructed for the translation of the Purcell swimmer. In the small-angle limit with $\theta_0=0$, the surfaces of non-equivalent strokes for net $x$ displacements behave identically to the squirmer but with three degrees of freedom instead of four because $\nabla \times \mathbf{M'}^{P}_{x}(\phi_{1},\phi_{2},\theta)$, Eq.~\eqref{pxdis}, is constant.  The surfaces for net $y$ displacement are however more complicated  since $\nabla \times \mathbf{M'}^{P}_{x}(\phi_{1},\phi_{2},\theta)$, Eq.~\eqref{pydis}, is not constant. In this case
\begin{equation}
\partial\mathbi{M}^{P}_{y}(\mathbf{S}^{p}) =  -\frac{2 \Delta \zeta}{81 \zeta_{\parallel}} \left(\begin{array}{c c c}
0 & -10 \theta & -(4 \phi_{1} + 5 \phi_{2}) \\
-10 \theta & 0& (5 \phi_{1} + 4 \phi_{2}) \\
(4 \phi_{1} + 5 \phi_{2}) &-(5 \phi_{1} + 4 \phi_{2})&0  \\
\end{array} \right),
\end{equation}
and the vector $\mathbf{v}$, which is perpendicular to $\partial\mathbi{M}^{P}_{y}\cdot\partial \mathbf{S}^{P}/ \partial t$,  can be written generally as
\begin{eqnarray}
\mathbf{v}^{P} &=& b(\mathbf{S}^{P}) \frac{\partial \mathbf{S}^{P}}{\partial t}   \\
&&+ c(\mathbf{S}^{P})\left(\begin{array}{c c c}
0 & -10 \theta(5 \phi_{1} + 4 \phi_{2})^2 & (4 \phi_{1} + 5 \phi_{2}) (5 \phi_{1} + 4 \phi_{2})^2 \\
-10 \theta (4 \phi_{1} + 5 \phi_{2})^2 & 0& (5 \phi_{1} + 4 \phi_{2}) (4 \phi_{1} + 5 \phi_{2})^2 \\
100 \theta^2(4 \phi_{1} + 5 \phi_{2}) & 100 \theta^2 (5 \phi_{1} + 4 \phi_{2})&0  \\
\end{array} \right) \cdot \frac{\partial \mathbf{S}^{P}}{\partial t}, \notag
\end{eqnarray}
where $\mathbf{S}^{P} = \{\phi_1(s,t),\phi_2(s,t),\theta(s,t)\}$. Hence the partial differential equation for the surfaces of non-equivalent strokes in $y$ is
\begin{equation}
\frac{\partial \mathbf{S}^{P}}{\partial s} = a(\mathbf{S}^{P}) \partial\mathbi{M}^{P}_{y}(\mathbf{S}^{p}) \cdot \frac{\partial \mathbf{S}^{P}}{\partial t} +\mathbf{v}^{P}, 
\end{equation}
for arbitrary choices of $a(\mathbf{S}^{P})$, $b(\mathbf{S}^{P})$ and $c(\mathbf{S}^{P})$. There is no choice of  $a(\mathbf{S}^{P})$, $b(\mathbf{S}^{P})$ and $c(\mathbf{S}^{P})$ and coordinate transform which can reduce these equations to the Cauchy–Riemann equations. This means that simple planes are not a surface of non-equivalent strokes for the displacement in $y$ and so planes will not, in general, contain every possible displacement that the swimmer can generate. This partial differential equation can be solved numerically to find the surfaces for a specific choice of $a(\mathbf{S}^{P})$, $b(\mathbf{S}^{P})$ and $c(\mathbf{S}^{P})$. Sections of three such surfaces for $a(\mathbf{S})=1$, $\mathbf{v}=0$ are shown in Fig.~\ref{fig:Purcell_ss}. These surfaces have been coloured according to the net displacement in $y$ the corresponding loop generates. Equivalent strokes between the surfaces can therefore be easily identified by looking for loops which sit within the same color. We note that if any stroke is taken backwards it generates a displacement of the same size but opposite sign.

\begin{figure}
\centering
\includegraphics[width=0.6\textwidth]{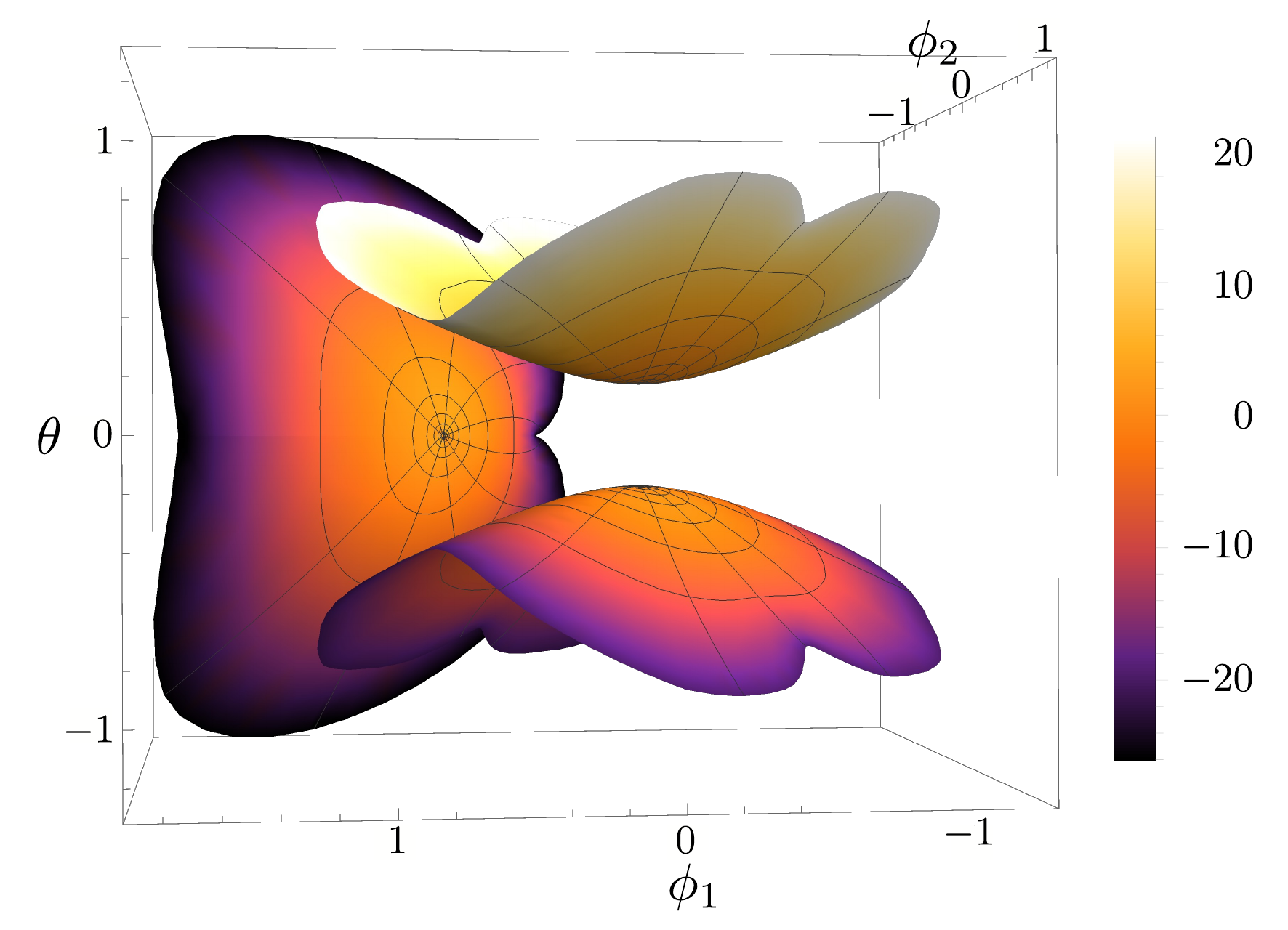}
\caption{{ Sections of three surfaces of non-equivalent strokes for the $y$ displacement of the Purcell swimmer. All these surfaces were formed with $a(\mathbf{S})=1$, $\mathbf{v}=0$ and the initial strokes were $\{\phi_1,\phi_2,\theta\}=\{0.01 \cos(t), 0.01 \sin(t), 0.5\}$, $\{0.01 \cos(t), 0.01 \sin(t), -0.5\}$ , $\{0.01 \cos(t)/\sqrt{2}, 0.01 \cos(t)/\sqrt{2}, 0.01 \sin(t)\}$. The color of each surface reflects the displacement in $y$ generated from each loop. } } 
\label{fig:Purcell_ss}
\end{figure} 

}
\section{Steps to design the displacement of a viscous swimmer} \label{sec:design}

In the above sections, we developed new techniques to treat non-commuting variables and the visualisation of phase space. These methods can be used to help design  unconstrained viscous swimmers for specific displacements. Here we outline one method to do this { and demonstrate it in full on a simple example}.

\subsection{General procedure}

Consider a viscous swimmer, which we want to travel  a total displacement $L$ along a specific axis after each stroke. The swimmer can deform its body using $N$ different modes and moves in three dimensions. From our analysis, we know that if the displacement $L$ is possible, there is an infinite number of `strokes' which could create it. The space of strokes that produce the displacement $L$ can be found through the following steps:
\begin{itemize}
\item[1.] Calculate the full displacement field, $\mathbi{M}(\mathbf{l},\mathbf{x})$, of the swimmer (Sec.~\ref{sec:dis1}).
\item[2.] Embed the field, $\mathbi{M}(\mathbf{l},\mathbf{x}) \to\mathbi{M}'(\mathbf{l}')$, to overcome non-commuting variables (Sec.~\ref{sec:commute}).
\item[3.]  Apply Stokes theorem to the field $\mathbi{M}'(\mathbf{l}')$  to find $\partial\mathbi{M}'(\mathbf{l}')$ (Sec.~\ref{sec:dis3}).
\item[4.] Use $\partial\mathbi{M}'(\mathbf{l}')$ to  obtain surfaces of non-equivalent strokes (Eq.~\ref{surf}, Sec.~\ref{sec:dis4}).
\item[5.] Identify a stroke of displacement $L$, if possible, in this set using Eq.~\eqref{cdx}.
\item[6.] Determine the equivalent strokes through symmetries or other solutions  (Sec.~\ref{sec:dis5}).
\end{itemize}
If only one stroke  producing the displacement $L$ is needed, this process can be stopped at step 5. Note that it is possible that no  single stroke can lead to a displacement $L$ in step 5. In that case, we can consider loops of displacement $L/n$, where $n$ is a positive integer;  these loops will then produce the displacement $L$ after  $n$ strokes. Also, the procedure outlined above will  work for a general swimmer though of course simpler methods could exist for specific systems.

{ 
\subsection{Example: Squirmer near a free interface} \label{sec:squirmint}

This process can be demonstrated in full using the example an axisymmetric squirmer near a free surface (Fig.~\ref{fig:squirmwall}). Though this swimmer does not generate any off axis rotation, the presence of the free surface means the swimmer's field will depend on the distance between the swimmer and the interface at all separations. This swimmer is therefore not isolated and its displacement cannot be found with the minimal perturbation coordinates treatment of Eq.~\eqref{Lie}. Here we will look for strokes which generate a net displacement of $L$ and explicitly go through each of the above steps.

\subsubsection{Calculate the displacement field}

Similar to the Purcell and squirmer examples above, to find the displacement field we need to perform a force balance on the squirmer in the presence of the interface. For this purpose we assume that the squirmer is very close to the interface, such that lubrication stresses dominate, and that the free interface is flat (Fig.~\ref{fig:squirmwall}). We will denote the clearance between the free interface and the squirmer $h$, the radius of the squirmer $1$ and consider the surface deformations
\begin{eqnarray}
R &=& 1 + \epsilon   \alpha_{2}(t) \cos\theta_{0} , \\
\theta &=& \theta_{0} +  \epsilon   \beta_{1}(t) \sin\theta_{0}.
\end{eqnarray}
The viscous hydrodynamics of a swimmer near a flat free interface is equivalent to that of the swimmer and a mirror swimmer on the other side of the interface \citep{Kim2005}. The force on the squirmer, from rigid body motion and the surface deformations, can therefore be determined through the hydrodynamic interactions of two spheres with prescribed surface velocities. Hence the hydrodynamic drag on the squirmer from motion perpendicular to the interface is the same as the drag on two approaching spheres and, in the lubrication limit $h \ll 1$, is given by \citep{Kim2005}
\begin{equation}
F_{H} = -\frac{3 \pi \mu}{2 h} \frac{d h}{d t}.
\end{equation}
This stress originates from the region on the squirmer that is closest to the interface and is caused by the large gradients in the velocity needed to meet the incompressible flow condition.

\begin{figure}
\centering
\includegraphics[width=0.6\textwidth]{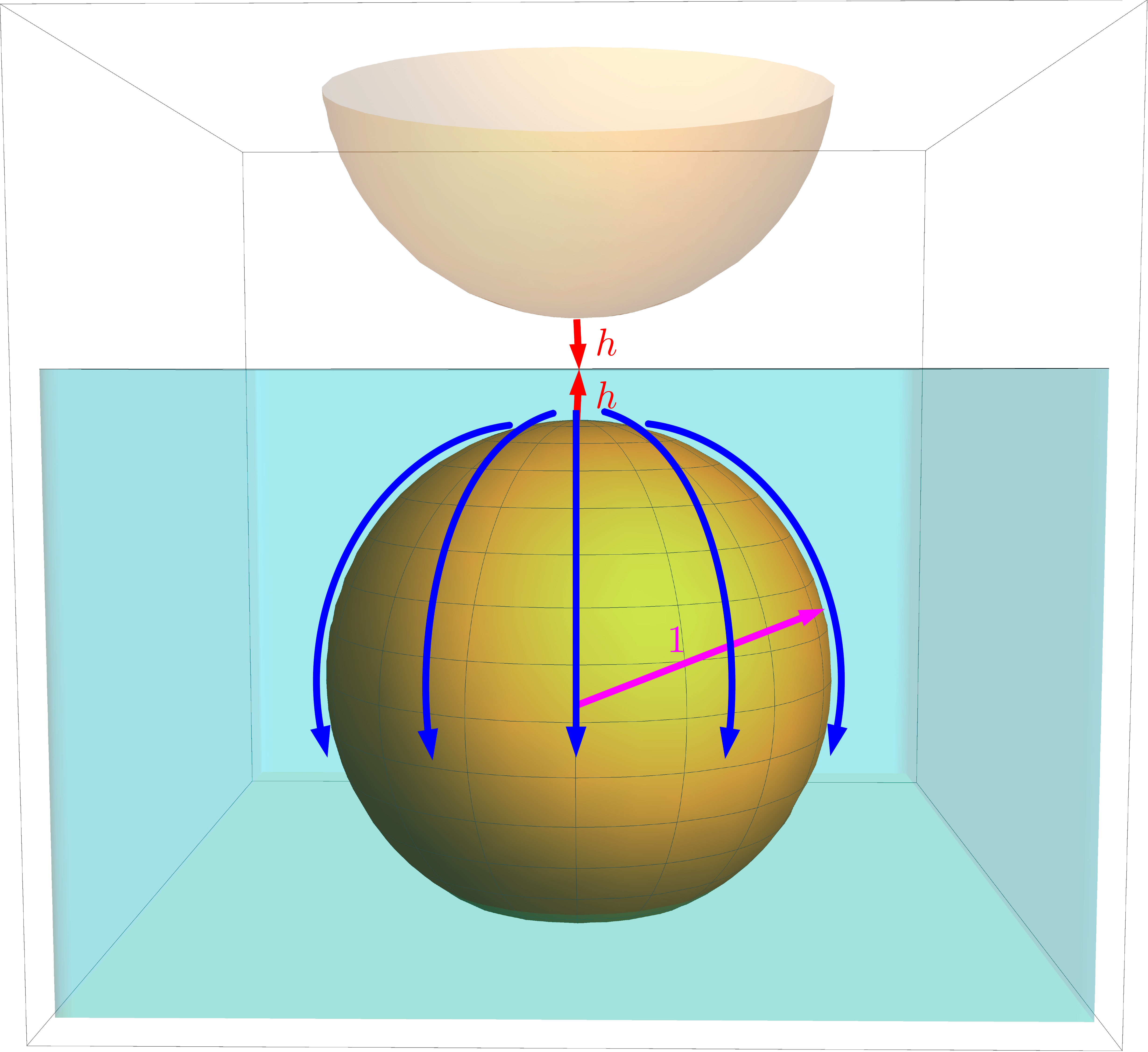}
\caption{{ Diagram depicting the configuration of the squirmer approaching the free surface. The squirmer has radius $1$ and the separation between the surface and the swimmer is $h$. The physics of a squirmer near a flat interface is the same as a squirmer approaching a mirror one with separation $2h$. }} 
\label{fig:squirmwall}
\end{figure} 

Similarly the surface velocity from the $\alpha_2(t)$ deformation mode, also generates lubrication flows. This is because around $\theta_0 =0$ (the point closest to the interface)  the surface velocity of the squirmer from this deformation mode is $\epsilon  d \alpha_{2} / d t$ and is directed towards the interface. Hence the flow in this nearly touching region for this mode is identical to that of approaching spheres with velocity $\epsilon  d \alpha_{2} / d t$. Since the flow in this region dominates the stress on the body \citep{Kim2005} the hydrodynamic force on the body from this mode is
\begin{equation}
F_{\alpha} = -\frac{3 \pi \mu \epsilon}{2 h}  \frac{d \alpha_{2}}{d t}.
\end{equation}
Unlike the rigid body motion and the flow from $\alpha_2$, the flows generated by the $\beta_{1}(t)$ deformation mode is not lubricating \citep{ISHIKAWA2006} and so the force on the lubrication scale $h \ll 1$ is roughly constant. For this example we write it as proportional to the free space squirmer velocity generated by this mode and given by
\begin{equation}
F_{\beta} = -4 \pi \mu \epsilon  k \frac{d \beta_{1}}{d t},
\end{equation}
where $k$ is a positive constant. 

These deformation forces can be balanced with the rigid body motion force to find
\begin{equation}
\frac{d h}{d t} =-  \frac{d \alpha_{2}}{d t}  -\frac{8 \epsilon k }{3} h(t)  \frac{d \beta_{1}}{d t} = \mathbf{M}^{si} \cdot \left\{\frac{d \alpha_{2}}{d t},  \frac{d \beta_{1}}{d t}\right\} \label{sqint}
\end{equation}
where $\mathbf{M}^{si} = - \epsilon \{1, 8 k h/3 \}$. This field has same structure as in Eq.~\eqref{demo} and has the exact solution
\begin{equation}
h(t) = h(0) \exp\left(-\frac{ 8\epsilon k}{3} [\beta_1(t)-\beta_1(0)] \right)-  \exp\left(-\frac{ 8\epsilon k}{3} \beta_1(t) \right) \int_{0}^{t} \exp\left(\frac{ 8\epsilon k}{3} \beta_1(t') \right)  \frac{d \alpha_{2}(t')}{d t'} \,d t'.
\end{equation}
Though we have this solution, the presence of the exponentials make it non-trivial to identify the strokes which generate displacements $L$. We will therefore show how we can do this by embedding the field and exploring the space of equivalent and non-equivalent strokes.
 
\subsubsection{Embed the field}

After identifying the displacement field the next step in finding the strokes of displacement $L$ is to embed the field such that non-commuting variables are now treated as prescribed paths. This can be achieved by simply adding zeros to the end of $\mathbf{M}^{si}$ to account for each non-commuting variable. Hence, in the case of the squirmer the embedded field is $\mathbf{M'}^{si} =   - \epsilon \{1, 8 k h/3, 0 \}$ and the net displacement can be written as
\begin{equation}
\Delta h = \oint_{\partial V}  \mathbf{M}^{si} \cdot \{\,d \alpha_2, \,d \beta_1\} = \oint_{\partial V'}  \mathbf{M'}^{si} \cdot \{\,d \alpha_2, \,d \beta_1, \,d h\},
\end{equation}
where $\partial V$ is the loop in $\alpha_2,\beta_1$, and $\partial V' = \partial V + k$ is the stroke embedded into the higher plane plus a path $k$ which closes the loop but adds nothing to the displacement.

\subsubsection{Apply Stokes theorem}

In the embedded representation we can now apply the generalised stokes theorem to find
\begin{equation}
\Delta h =  \oint_{\partial V'}  \mathbf{M'}^{si} \cdot \{\,d \alpha_2, \,d \beta_1, \,d h\} = \iint_{V'}  \partial\mathbf{M'}^{si} :  \left(d \boldsymbol{l}^{si} \wedge d \boldsymbol{l}^{si}\right),
\end{equation}
where $V'$ is a surface bounded by the stroke and
\begin{eqnarray}
\partial\mathbf{M'}^{si} &=& \frac{4 k \epsilon}{3} \left(\begin{array}{c c c }
0 & 0 & 0  \\
0& 0& 1\\
0& -1&0  
\end{array} \right),\\
d \boldsymbol{l}^{si} \wedge d \boldsymbol{l}^{si} &=& \left(\begin{array}{c c c}
0 & d\alpha_{2}\wedge d \beta_{1} & d\alpha_{2}\wedge d h    \\
-d\alpha_{2}\wedge d \beta_{1}  & 0 &d\beta_{1}\wedge d h \\
-d\alpha_{2}\wedge d h & d\beta_{1}\wedge d h & 0 
\end{array} \right).
\end{eqnarray}

\subsubsection{Determine the surfaces of non-equivalent strokes}

Next we take the results from the generalised Stokes theorem and substitute it into Eq.~\eqref{surf} to look for surfaces on non-equivalent strokes. For the squirmer by an interface these equations become
\begin{eqnarray}
\frac{\partial \alpha_{2}(s,t)}{\partial s} &=& 0, \\
\frac{\partial \beta_{1}(s,t)}{\partial s} &=& \frac{4 k \epsilon}{3} \frac{\partial h(s,t)}{\partial t},   \\
\frac{\partial h(s,t)}{\partial s} &=& -\frac{4 k \epsilon}{3} \frac{\partial \beta_{1}(s,t)}{\partial t},
\end{eqnarray}
which has the general solution 
\begin{equation}
\mathbf{S}(s,t) = \{\alpha_{2}(s,t), \beta_{1}(s,t), h(s,t)\} = \left\{ f(t), \Re\left[g\left(\frac{3}{4 k \epsilon}s + i t\right) \right], \Im\left[g\left(\frac{3}{4 k \epsilon}s + i t\right) \right]\right\},
\end{equation}
where $g(z) $ is an analytic function and $f(t)$ is a periodic function in $t$. Since $f(t)$ is arbitrary and does not depend on $s$, distortions in $\alpha_{2}$ do not contribute to the net displacement which only depends on the area enclosed by the stroke within the $\beta_{1}$-$h$ plane in the embedded space. This is reflected by the dependence on the analytic function $g(z)$, similarly to a squirmer in free space.

\subsubsection{Identify desired displacement}

The general solution for the surfaces of non-equivalent strokes indicate that the displacement from any stroke solely depends on the area enclosed by the loop when projected onto the $\beta_{1}$-$h$ plane. This is captured by the dependence of the solution on the analytic function $g(z)$, with the different choices of $g(z)$ representing all the different parametrisation of this plane possible. Swimming strokes which generates net displacements of size $L$, for a given choice of $g(z)$, are the strokes which enclose an area $3 L/ 8 k \epsilon$ within it. For example consider the parametrisation $g(z) = \cosh(z)$ with $f(t)=0$. In which case the surface of non-equivalent strokes is given by
\begin{equation}
\mathbf{S}_{ex}(s,t) =  \left\{ 0,\cosh \left(\frac{3 s}{4 k \epsilon} \right)\cos(t), \sinh \left(\frac{3 s}{4 k \epsilon} \right)\sin(t)\right\},
\end{equation}
and the displacement from each $s$ is given by
\begin{eqnarray}
\Delta h &=& \iint_V'  \partial\mathbf{M'}^{si} :  \left(\frac{d \boldsymbol{S}_{ex}}{d t} \wedge \frac{d \boldsymbol{S}_{ex}}{d s}\right) \,ds \,dt \notag \\
 &=&\frac{8 k \epsilon}{3} \int_{0}^{2\pi} \,d t \int_{0}^{s} \,d s' \frac{-3}{8 k \epsilon} \left[ \cos( 2 t) - \cosh\left(\frac{3 s'}{2 k \epsilon} \right) \right]  \notag \\
 &=& \frac{4 \pi k \epsilon}{3} \sinh \left(\frac{3 s}{2 k \epsilon} \right),
\end{eqnarray}
which provides us with the displacement $L$ when $s = 2 k \epsilon \mbox{ arcsinh}( 3L/4 k \pi \epsilon )/3$. 

\subsubsection{Determine equivalent strokes}

In the above section we found that any loop in a $\beta_{1}$-$h$ plane that encloses an area of $3 L/ 8 k \epsilon$ generates a displacement of size $L$. Hence all these strokes are equivalent. Furthermore, we found that $\alpha_{2}$ does not modify the displacement in the embedded representation and so any stroke that encloses an area $3 L/ 8 k \epsilon$ in the $\beta_{1}$-$h$ plane with arbitrary $\alpha_{2}(t)$ also generates the net displacement $L$. With this we have identified the all the equivalent strokes with net displacement $L$ and can now choose a stroke from this set that best suits our purposes.

In the case of the squirmer by an interface we can further restrict this set of equivalent strokes to isolate the strokes that also satisfy Eq.~\eqref{sqint}. Since the displacement of this squirmier only depends on the stroke taken in $\beta_{1}$-$h$ we can rearrange the original equation to solve for $\alpha_{2}$.
This gives us
\begin{equation}
 \alpha_{2}(t) - \alpha_{2}(0)  =-h(t)+h(0)    -\frac{8 \epsilon k }{3} \int_{0}^{t} h(t')  \frac{d \beta_{1}(t')}{d t'} \,d t', \label{alpha}
\end{equation}
which we can solve for any prescription of $h(t)$ and $\beta_1(t)$. In addition to the above we also require that $ \alpha_{2}(t)$ and $\beta_{1}(t)$ are periodic over the stroke $\partial V$ while $h(t)$ is arbitrary. This is because in the embedded space we can always add an additional path $k$ which holds $\alpha_2$ and $\beta_1$ constant but returns $h(t)$ to its original value without changing the net displacement. Hence if the stroke has a period of $2 \pi$ this periodicity condition becomes
\begin{equation}
L = h(2 \pi)-h(0) = -\frac{8 \epsilon k }{3} \int_{0}^{2 \pi} h(t')  \frac{d \beta_{1}(t')}{d t'},
\end{equation}
where we have used that we are considering strokes that displace $L$. If we expand $\beta_{1}(t)$ in terms of the Fourier series
\begin{equation}
\beta_{1}(t) = A_0 + \sum_{n=1}^{\infty} A_n \cos(n t) + B_n \sin(n t), \label{beta1}
\end{equation}
the above relationship can be written as
\begin{equation}
-\frac{ 3 L }{8 \epsilon k} = \sum_{n=1}^{\infty} n\left[ B_n \int_{0}^{2 \pi} h(t')\cos(n t') \,d t'  -n A_{n} \int_{0}^{2 \pi} h(t')\sin(n t') \,d t'  \right], \label{beta2}
\end{equation}
and can be used to determine one of the coefficients in $\beta_1$. Hence the strokes with displacement $L$ and satisfy Eq.~\eqref{sqint} are loops  which enclose an area of $3 L/ 8 k \epsilon$ in the $\beta_{1}$-$h$ plane, satisfy Eq.~\eqref{beta2}  and have $\alpha_{1}$ specified by Eq.~\eqref{alpha} over $\partial V$. These can be formed by selecting the path in $h(t)$ desired, using Eq.~\eqref{beta2} to determine one of the coefficients in $\beta_{1}$, choosing the others and then finally substituting these paths into Eq.~\eqref{alpha}. Hence, through this method we can identify all the strokes which generate a displacement $L$ and satisfy Eq.~\eqref{sqint}.
}
 \section{Conclusion} \label{sec:con}
 
In this paper we propose a method to identify, and explore the displacement of arbitrary viscous swimmers by combining geometric swimming techniques and Stokes theorem. Typically, issues with non-commuting variables and visualisation would prevent the direct application to arbitrary swimmers.   We developed novel methods to overcome these issues and thereby showed that the set of possible displacements achieved by any viscous microswimmer in any environment can always be visualised using a single surface. 
 
 We first showed that variables that do not commute can be treated as parametrised paths by embedding the path integral representation into a higher dimensional configuration space.  This higher-dimensional treatment inherently captures the displacement of the swimmer in all possible situations, is mathematically exact and can be performed by hand but takes the configuration paths specific to a problem as an input. It also allows the generalized Stokes theorem to be applied directly to every situation.
 
 We then showed that if the generalized Stokes theorem can be applied all the net displacements possible can be visualised on special surfaces called surfaces of non-equivalent strokes. This is because the generalized Stokes theorem always gives the net displacement in terms of the flux of a divergent-free field through a surface. This divergence free property implies that an infinite set of loops must exist for any net displacement and  that no one swimming stroke will generate the maximum displacement without additional conditions. We used the properties of this field to create surfaces of strokes in the configuration space that contain all possible net displacements for the swimmer in every situation. This identification and construction procedure has never been done previously to our knowledge. The displacements available to a specific physical system with constraints occupy a continuous region on these surfaces. Loops on different surfaces which generate the same displacement were then identified by considering the change in net displacement with the loop parametrisation. In the special case of a constant field, we also showed that all loops can be projected onto a single plane and suggest a method to create equivalent loops from an initial one. 
 
{ Finally we draw these methods together to describe a general procedure to explore and design the motion of a general swimmer. This involves the combination of the embedding of the path integral to overcome non-commuting strokes and the surface construction techniques to determine a set of non-equivalent paths. This procedure can be applied to any swimmer in any environment and is demonstrated on a squirmer near a free surface.} 
 
 The methodology developed in this paper considered an idealised system  and treated all possible environmental cases simultaneously.  However, we are typically interested in swimmers in a given environment with  constraints like no-external force, constant volume, or constant surface area. Hence the methods described could be applied more broadly if a generic procedure to identify and include these limits could be developed. This would also reveal how constraints would affect our results on the degeneracy of equivalent stokes  and has been identified as important future work. Additionally it would be good to see if these surface creation ideas can be of use in the optimisation of swimmers, through the addition of cost metrics like in \citet{Ramasamy2017, Ramasamy2019}. Finally the discussion also raises the question about what strokes would be equivalent to those commonly seen in nature.

\acknowledgements

This project has received funding from the European Research Council (ERC) under the European Union's Horizon 2020 research and innovation programme  (grant agreement 682754 to EL), and the Australian Research Council (ARC) under the Discovery Early Career Research Award (grant agreement DE200100168 to LK). LK was also supported by a Macquarie University new staff grant.

 Declaration of Interests: The authors report no conflict of interest.
\appendix

\section{Calculating the exterior derivatives} \label{sec:extior} 

In this appendix we outline how to calculate of the exterior derivative of the vector-like objects called 1-form to produce the matrix-like structures called 2-forms. A thorough introduction to exterior derivatives can be found in \citet{Crane2018}. 
 
A 1-form can generally be written as
\begin{equation}
\mathbf{w} \equiv w_{i} d x^{i},
\end{equation}
where $w_{i}$ is the component of the 1-form in the direction $d x^{i}$ and we have used the Einstein summation convention. This representation is coordinate dependant and so the $w_{i}$ change in different coordinates. The wedge product of two 1-forms is the antisymmetric tensor product and produces a 2-form.  Wedge products therefore have the properties $\mathbf{w}\wedge\mathbf{w} = 0$ and $ \mathbf{w} \wedge \mathbf{v}  = - \mathbf{v} \wedge \mathbf{w}$. The value of the wedge product is given by
\begin{equation}
\mathbf{w} \wedge \mathbf{v} \equiv  (w_{i} d x^{i})\wedge (v_{j} d x^{j})= \frac{1}{2}\left(w_{i} v_{j} -w_{j} v_{i}\right) dx^{i} \wedge d x^{j} .
\end{equation}
It is worth noting that the coefficients of any two form can always be represented by a skew symmetric matrix like
\begin{equation}
w_{i} v_{j} -w_{j} v_{i} \equiv \left(\begin{array}{c c c c }
0 & \left(w_{1} v_{2} -w_{2} v_{1}\right) & \left(w_{1} v_{3} -w_{3} v_{1}\right)  &  \cdots \\
-\left(w_{1} v_{2} -w_{2} v_{1}\right)  & 0 & \left(w_{2} v_{3} -w_{3} v_{2}\right) &  \\
-\left(w_{1} v_{3} -w_{3} v_{1}\right) & -\left(w_{2} v_{3} -w_{3} v_{2}\right)& 0 & \\
\vdots & & &\ddots
\end{array} \right).
\end{equation}
We note that higher order forms can be constructed by taking the wedge product of lower forms.

The above definitions now allows us to define the exterior derivative $d$. Similarly to the 1-form being like a vector, the exterior derivative is akin to an antisymmetric gradient operation. The exterior derivative applied to a scalar is
\begin{equation}
d \phi = \frac{\partial \phi}{\partial x^{i}} d x^{i}.
\end{equation}
while the exterior derivative applied to a 1-form is
\begin{eqnarray}
d \mathbf{w} &\equiv &  d(w_{i} d x^{i}) = d(w_{i}) \wedge d x^{i} +  w_{i} \wedge d(d x^{i})  \notag\\
&=& \left(\frac{\partial w_{i}}{\partial x^{j}} d x^{j} \right) \wedge d x^{i} = \frac{1}{2} \left( \frac{\partial w_{i}}{\partial x^{j}} -\frac{\partial w_{j}}{\partial x^{i}}  \right) d x^{j} \wedge d x^{i},
\end{eqnarray}
where $d^{2} w =0$ for any $w$ because the operation is antisymmetric. 
The definition of the exterior derivative also provides us with the means to write down the generalized Stokes theorem,
 \begin{equation}
\int_{\partial V} w =\int_{ V} dw, 
\end{equation} 
where $w$ is a N-form, $V$ is the $N+1$-volume and $\partial V$ is the boundaries of said volume. In our paper, we are only interested in at most 2-forms and so the relationship becomes a surface and the edge of the surface.

\section{Properties of skew symmetric matrices} \label{sec:skew}  
 
 Any $N\times N$ skew symmetric matrix, $\mathbf{A}$, has complex eigenvalues that occur in pairs, i.e.
\begin{eqnarray}
\begin{array}{c c}
\pm i \lambda_{1}, \pm i \lambda_{2}, \cdots \pm i \lambda_{N/2} & \mbox{if $N$ is even}, \\
0, \pm i \lambda_{1}, \pm i \lambda_{2}, \cdots \pm i \lambda_{(N-1)/2} & \mbox{if $N$ is odd}, \\
\end{array}
\end{eqnarray}
where $\lambda_{1}, \lambda_{2}, \cdots$ are real positive numbers. As such no real transformation can diagonalise the system. However, there is always a rotation matrix $\mathbf{Q}$ such that
\begin{equation}
\mathbf{Q}^{T} \cdot \mathbf{A} \cdot \mathbf{Q} = \left(\begin{array}{c c c c c}
\begin{array}{c c}
0 & \lambda_{1} \\
- \lambda_{1} & 0
\end{array} & \mathbf{0} & \cdots & \mathbf{0} & \cdots \\
\mathbf{0}  &  \begin{array}{c c}
0 & \lambda_{2} \\
- \lambda_{2} & 0
\end{array} & &  \mathbf{0} & \cdots \\
\vdots &  & \ddots & \vdots &  \\
\mathbf{0}  & \mathbf{0}  & \cdots & \begin{array}{c c}
0 & \lambda_{n} \\
- \lambda_{n} & 0
\end{array}  & \cdots \\
\vdots & \vdots & & \vdots & \ddots
\end{array} \right),
\end{equation}
where $\cdot^{T}$ is the matrix transpose.

 The flux through a surface element is therefore
\begin{equation}
A_{ij} d l^{i} \wedge d l^{j} = 2 \lambda_{1} d x^{1} \wedge d x^{2}  + 2 \lambda_{2} d x^{3} \wedge d x^{4}  + \cdots 2 \lambda_{n} d x^{2n-1} \wedge d x^{2n} + \cdots \label{rot}
\end{equation}
were $d x^{i} = Q_{ij} d  l^{j}$ is the rotated basis directors. This representation separates the flux into non-interacting components. For example the flux through a surface that sits purely in the $x^1$ and $x^2$ plane is the same as a plane that sits purely in $x^{1}$, $x^2$ and $x^3$ because $A_{ij}$ has no $A_{13}$ or $A_{23}$ components in this frame.  The rotated directors, $d x^{i}$, are therefore a natural representation for the calculation of the flux through surfaces.

The above discussion considered a constant skew symmetric matrix, $A_{ij}$. The displacement field, $\displaystyle \left( \frac{\partial M_{ij}}{\partial l^{k}} -\frac{\partial M_{ik}}{\partial l^{j}}  \right)$, however, will vary with the position in space. The eigenvalues, $\lambda_i$, the rotation matrix, $\mathbf{Q}$, and directors, $d x^{i}$, will thus also vary with the position in space. Furthermore, the directors do not typically prescribe a new coordinate system because the displacement field is non-conservative.

\bibliographystyle{jfm}
\bibliography{library}
\end{document}